\documentclass{jfm}
\usepackage{graphicx}
\usepackage{amsmath}
\usepackage{subfigure}
\usepackage{psfrag}
\usepackage{epstopdf, epsfig}
\usepackage{hyperref}
\usepackage{ar}
\hypersetup{
	bookmarksopen=true,         % show bookmarks bar?
	unicode=false,          % non-Latin characters in Acrobat’s bookmarks
	pdftoolbar=true,        % show Acrobat’s toolbar?
	pdfmenubar=true,        % show Acrobat’s menu?
	pdffitwindow=false,     % window fit to page when opened
	pdfstartview={FitH},    % fits the width of the page to the window
	pdftitle={My title},    % title
	pdfauthor={Author},     % author
	pdfsubject={Subject},   % subject of the document
	pdfcreator={Creator},   % creator of the document
	pdfproducer={Producer}, % producer of the document
	pdfkeywords={keyword1, key2, key3}, % list of keywords
	pdfnewwindow=true,      % links in new PDF window
	colorlinks=true,       % false: boxed links; true: colored links
	linkcolor=blue,          % color of internal links (change box color with linkbordercolor)
	citecolor=blue,        % color of links to bibliography
	filecolor=blue,        % color of file links
	urlcolor=blue          % color of external links
}

\newcommand{\diff}{{\mathrm d}}

\shorttitle{DNS of conical shock/boundary layer interaction}
\shortauthor{Zuo F. Y., Memmolo A., Huang G. P. and Pirozzoli S.}

\title{Direct numerical simulation of conical shock wave/turbulent boundary layer interaction}

\author{Feng-Yuan Zuo\aff{1,2}
  \corresp{\email{zuofy@nuaa.edu.cn}},
  Antonio Memmolo\aff{2},
  Guo-ping Huang\aff{1},
   \and Sergio Pirozzoli\aff{2}
   \corresp{\email{sergio.pirozzoli@uniroma1.it}}}

\affiliation{\aff{1}College of Energy and Power Engineering, Nanjing University of Aeronautics and Astronautics, Nanjing, 210016, China
\aff{2} Sapienza Universit\`a di Roma, Dipartimento di Meccanica e Aerospaziale, Via Eudossiana 18, 00184 Roma, Italy}

\date{\today}
\begin{document}

\maketitle

\begin{abstract}
Direct numerical simulation of the Navier-Stokes equations
is carried out to investigate the interaction of a conical shock wave with
a turbulent boundary layer developing over a flat plate at free-stream
Mach number $M_\infty = 2.05$ and Reynolds number $\Rey_\theta \approx 630$,
based on the upstream boundary layer momentum thickness.
The shock is generated by a circular cone
with half opening angle $\theta _c = 25^\circ$.
As found in experiments, the wall pressure exhibits a distinctive N-wave signature,
with a sharp peak right past the precursor shock generated at the cone apex, 
followed by an extended zone with favourable 
pressure gradient, and terminated by the trailing shock
associated with recompression in the wake of the cone.
The boundary layer behavior is strongly affected by the imposed pressure gradient,
with streaks which are locally suppressed in adverse pressure gradient (APG) zones, and which
suddenly reform in the downstream region with favourable pressure gradient (FPG). 
Three-dimensional mean flow separation is only observed in the 
first APG region associated with formation of a horseshoe vortex, whereas the second APG region 
features an incipient detachment state, with scattered spots of
instantaneous reversed flow. 
As found in canonical two-dimensional wedge-generated 
shock/boundary layer interactions, different amplification of the turbulent stress components 
is observed through the interacting shock system, with approach to isotropic state
in APG regions, and to a two-component anisotropic state in FPG.
\end{abstract}


\section{Introduction}\label{sec:introduction}

Shock wave/turbulent boundary layer interactions (SBLI) 
occur whenever a shock sweeps across the boundary layer developing on a wall surface,
and as a consequence they have great interest
in aeronautics and aerospace engineering.
SBLIs are found in several situations of practical importance, such as wing-fuselage and tail-fuselage junctions of an aircraft,
helicopter blades, supersonic intakes, over-expanded nozzles, launch vehicles during the ascent phase, etc.
Typically, these phenomena have a significant drawback on aerodynamic performance,
yielding loss of efficiency of aerodynamic surfaces, unwanted wall pressure fluctuations
possibly leading to vibration and fatigue of structural components,
and to localized heat transfer loads, especially in presence of flow separation~\citep{Dolling2001-Fifty,Smits-2006-Turbulent,Babinsky2011-Shock}.
A large number of experimental and numerical studies,
along with the concurrent development of flow control techniques, have been carried out in past decades,
but nevertheless the unsteady features and the turbulence amplification
mechanisms conveyed by SBLI are not fully understood, remaining a challenging state-of-the-art research problem.

Most studies of SBLI have been carried out in idealized settings involving geometrically
two-dimensional configurations, and/or using simplified modeling approaches~\citep{Delery2009-physical,DeBonis2012-Assessment,Morgan2013-physics}.
As a common conclusion, these investigations have proved
that turbulence models have the ability to capture the mean flow features
such as the pressure loads and the interaction length scales, but obviously they cannot
capture the unsteady features and correctly predict flow separation.
The most relevant studies on two-dimensional SBLI have been carried out experimentally
or numerically through high-fidelity simulations
~\citep{Adams2000-Direct,Pirozzoli2006-Direct,Smits-2006-Turbulent,Dupont2006-Space,Pirozzoli2011-Direct,Touber2011-Low-order,Hadjadj2012-Large-Eddy}.
The overall picture involves a boundary layer which develops under APG conditions,
with the effects of the shock that are felt upstream of the impingement as
a result of propagation of information in the subsonic part of the boundary layer.
In case the shock is strong enough,
separation takes place and pressure shows a plateau in the separation region,
with a typical low-frequency motion of the shock system and of the separation bubble.
Turbulence experiences amplification while traversing the shock, and it undergoes a relaxation
process as is proceeds past the interaction region.
However, SBLI occurring in flow conditions of practical relevance are 
almost invariably three-dimensional in nature.
Most often, shock waves are generated by finite-sized bodies,
and interact with turbulent boundary layers developing on solid surfaces.
\begin{figure}
  \centering 
   (a)\includegraphics[height=3.2cm,clip]{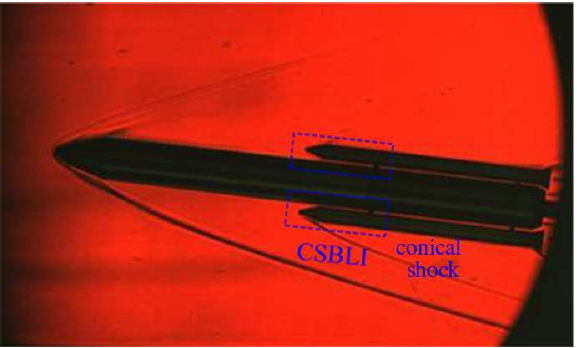}
   (b)\includegraphics[height=3.2cm,clip]{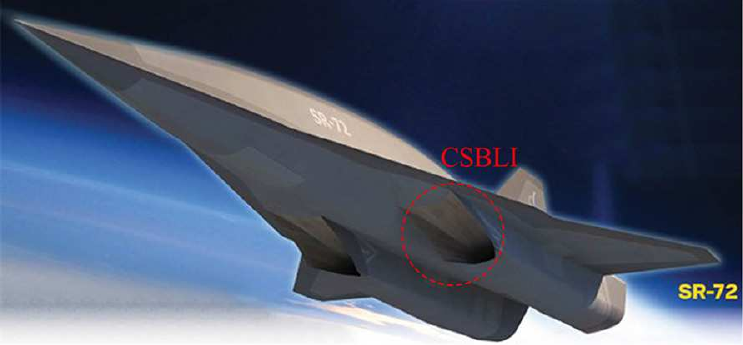} \\ \vskip1.em
   \caption{Conical SBLI in 
	launch vehicle during the ascent phase (a) and 
    in the air intake of the hypersonic SR-72 vehicle (b).}
	\label{fig:CSBLI_examples}
\end{figure}
\begin{figure}
	\centering{\includegraphics[width=12.0cm,clip]{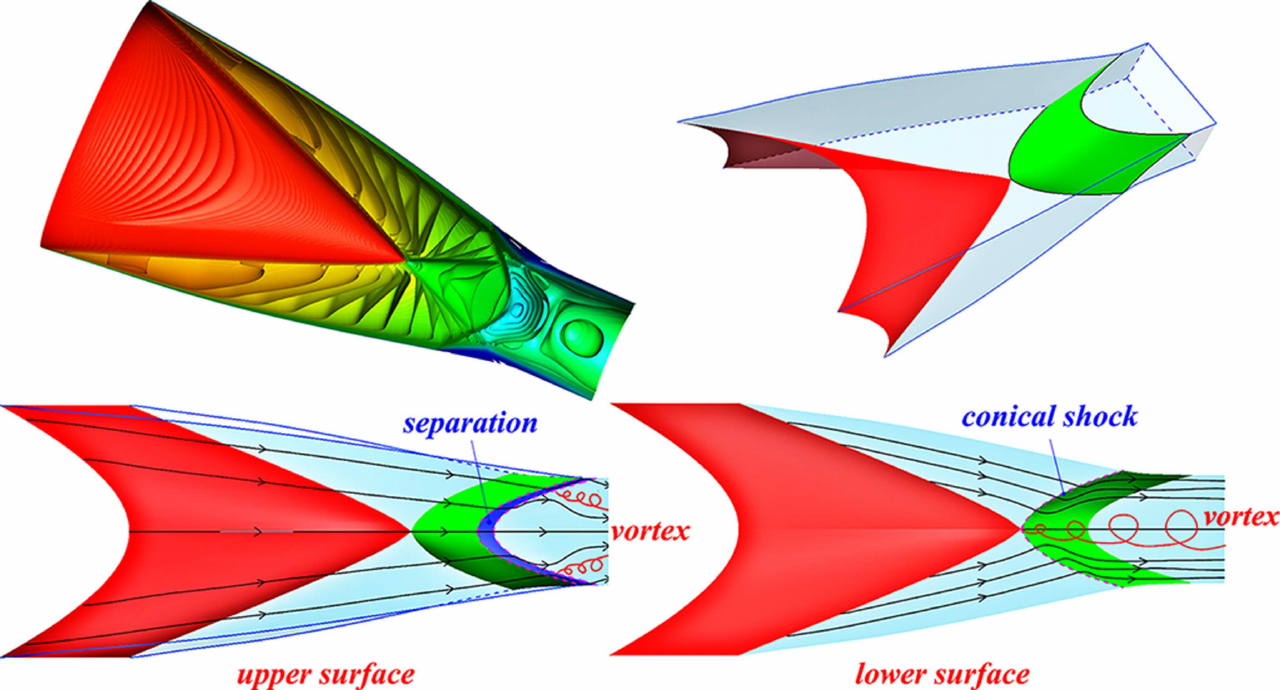}}
	\caption{Sketch of flow structure in wave-catcher inward-turning intake~\citep{Zuo2018-Numerical}.
	Red and green surfaces are conical shocks,
        the purple surface is the separation region, red lines are a sketch
        of vortices caused by CSBLI.}
	\label{fig:CSBLI-intake}
\end{figure}
Conical shocks in particular are frequently found in high-speed vehicles,
both in external and internal flows, for which examples are provided in
figure~\ref{fig:CSBLI_examples}. 
In the supersonic ascent phase, multi-body launch vehicles feature conical 
shock boundary layer interactions (CSBLI) associated with reflection of the 
boosters-generated shocks with the main body of the launcher (see panel a, from~\citep{Sziroczak2016-A}).
CSBLI also occur in internal flows as in the hypersonic wave-catcher (inward-turning) intake
adopted on the SR-72 vehicle~\citep{Zuo2016-Investigation,Zuo2018-Numerical}, sketched in panel (b).
As shown in figure~\ref{fig:CSBLI-intake}, the initially conical shock wave covers the intake edge,
and it reflects as another conical shock impinging on the upper surface.

\begin{figure}
	\centering{\includegraphics[width=7.0cm]{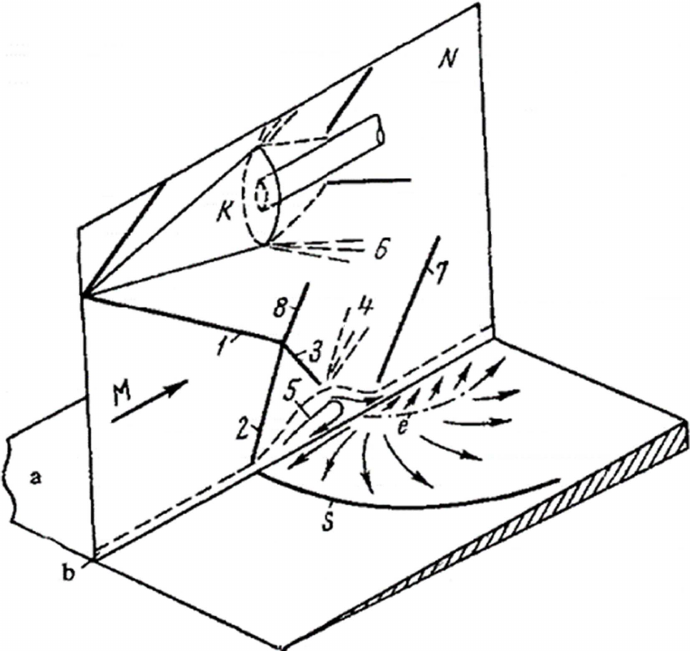}}% Images in 100% size
	\caption{Flow-field structure of strong CSBLI under separation condition~\citep{Panov1968-Interaction}.
	Numbered solid lines are conical shock traces;
	numbered dashed lines are rarefaction waves;
        $5$ is the separation region, delimited in the wall plane by
	$S$ (separation) and $e$ (reattachment).}
	\label{fig:CSBLI-sketch}
\end{figure}

The analysis of CSBLI is inherently more difficult than for planar SBLI, in that
the flow field past a conical shock is not uniform, and the wall pressure rise
is not uniform along the boundary-layer transverse direction, hence the resulting limiting wall streamlines are not parallel.
Moreover, non-uniformity of the imposed shear yields a variety of complex vortical structures which
interact and merge while becoming entrained in the main flow.
Greater challenge is also encountered in the numerical simulation of CSBLI, 
mainly because of slow convergence of the flow statistics in the absence of directions with
spatial homogeneity.
The leading features of CSBLI are sketched in figure~\ref{fig:CSBLI-sketch}, taken from \citet{Panov1968-Interaction}.
The shock generator K generates the incident conical shock 1,
and 2 is the three-dimensional separation shock associated with separation of the boundary layer.
The approaching flow that passes through shock 2 is deflected upward,
and after passing through shock 3 and rarefaction wave 4, the separated boundary layer is
directed at some angle toward the plate surface.
Shock 7 then arises to realign the outer flow to the wall-parallel direction.
The boundary layer, which separates along line S, reattaches along line e.
As in the case of the later analysis, the rarefaction wave $6$
emanating from the trailing edge of the shock-generating device can also
enter into play thus further complicating the analysis of the flow in the interaction region.
The overall phenomenon, with special reference to the separation region 5 is inherently three-dimensional in nature.

A mathematical analysis of CSBLI was carried out by~\citet{Migotsky1951-Three-Dimensional}
One of the main conclusions was that, even when regular reflection is possible
at the leading edge of the interaction zone, transition to Mach reflection shall occur 
somewhere along the spanwise direction.
\citet{Gai2000-Interaction} experimentally investigated the interaction of the shock wave produced by a conical
shock with a planar turbulent boundary layer, at free-stream Mach number 2. 
By varying the cone opening angles from $14^{\circ}$ to $30^{\circ}$, both attached and separated
conditions were achieved.
The results showed that the incident shock imposes a significant pressure gradient in the spanwise direction
yielding strong cross flow and formation of a horseshoe vortex whose signature was
evident in surface oil-flow visualizations, and whose size and strength gradually reduce away from the
symmetry plane. Notably, the pressure rise for incipient separation was found
to be less than for the two-dimensional case. 
\citet{Hale2015-Interaction} studied the impingement of a conical shock wave on a plane
turbulent boundary layer at Mach number 2.05, and gathered information by means of surface
oil flow, pressure-sensitive paint and particle image velocimetry techniques.
The experimental data suggested that the interaction causes locally two-dimensional separation
near the centerline, and three-dimensional separation away from this region, with
fluid propagating away from the centerline. 
Significant spanwise and streamwise expansion was observed right downstream of the 
interaction leading edge, unlike in equivalent two-dimensional wedge-generated SBLI.
The results further gave hints for instantaneous boundary layer separation, 
and showed that the interaction tends to suppress 
large-scale vortical structures in the incoming boundary layer.

Based on the survey of the state-of-the-art in CSBLI we believe that further
study is appropriate. In particular, we find that limited information about the detailed 
three-dimensional structure of the interaction has been gained through experiments,
and no high-fidelity computation (i.e. LES and/or DNS) has been reported so far.
Hence, we believe that a DNS-based analysis of CSBLI can help the research community 
in achieving improved physical understanding of the phenomenon, and to developed better
turbulence models for simplified prediction.
To make our analysis sufficiently general, we consider a relatively simple
geometrical set-up, whereby a fully developed turbulent boundary layer progresses on a flat solid surface,
and the shock wave generated by a circular cone with axis parallel to the wall is made to interact with it,
mimicking the experimental flow conditions of~\citet{Hale2015-Interaction}. 
The paper is organized as follows.
The numerical methodology is described in \S\ref{sec:computational_strategy};
the results are presented in \S\ref{sec:results_and_discussion},
covering a quantitative analysis of the mean and statistical properties
of the flow field. Concluding remarks are given in \S\ref{sec:conclusions}.

\section{Computational strategy}\label{sec:computational_strategy}

We solve the three-dimensional Navier-Stokes equations for a perfect Newtonian gas.
The molecular viscosity $\mu$ is assumed to obey Sutherland’s law,
and the thermal conductivity $k$ is related to $ \mu $ through
$k = {c_p}\mu /\Pr$,
where the molecular Prandtl number $Pr$ is assumed to be 0.72.
The Cartesian solver used for the present analysis was extensively
tested for compressible wall-bounded flows,
also in presence of impinging shocks~\citep{Pirozzoli2010-Direct,Pirozzoli2011-Direct}.
The main feature of the solver is the use of a conservative discretization of the convective fluxes
which combines sixth-order central non-dissipative discretization in smooth parts of the flow field
and seventh-order weighted-essentially non-oscillatory (WENO) discretization in shocked regions,
the switch between the two being controlled by a shock sensor
based on the ratio of local dilatation to vorticity modulus~\citep{Pirozzoli2011-Numerical}.
For the purpose of hybridization, critical grid nodes 
are preliminarily marked and then padded with four
nodes on both sides to ensure that the stencil of the underlying
non-dissipative scheme does not cross shocked zones.
Improved numerical stability for the central discretization in smooth
parts of the flow is achieved by splitting the convective derivatives as~\citep{Pirozzoli2010-Generalized}
\begingroup\makeatletter\def\f@size{9}\check@mathfonts
\begin{equation}
  \frac {\partial \rho u_j \varphi}{\partial x_j} = \frac 14 \, \frac {\partial \rho u_j \varphi}{\partial x_j} 
  + \frac 14 \, \left( u_j \, \frac {\partial \rho \varphi}{\partial x_j} 
             + \rho \, \frac {\partial u_j \varphi }{\partial x_j} 
             + \varphi \, \frac {\partial \rho u_j}{\partial x_j} \right)
  + \frac 14 \, \left( \rho u_j \frac {\partial \varphi}{\partial x_j} 
                        + \rho \varphi \, \frac {\partial u_j}{\partial x_j}  
                        + u_j \varphi \, \frac {\partial \rho}{\partial x_j} \right),
\label{eqn:kennedy}
\end{equation}\endgroup
\noindent where $ \varphi $ stands for a generic transported fluid property. The continuous derivative
operators are then replaced with central finite difference
approximations using a locally conservative formulation
\citep{Pirozzoli2010-Generalized}, which guarantees global conservation
of mass, momentum, and total energy through the telescopic property and
simplifies hybridization with the WENO algorithm.
An important property of the convective split form~\eqref{eqn:kennedy} 
is that it leads to kinetic energy preservation for
inviscid, incompressible flow, guaranteeing strong numerical stability
without reverting to upwinding or filtering.
The diffusive terms in the Navier-Stokes equations are also approximated
with sixth-order central differences after being expanded to Laplacian
form to guarantee finite molecular dissipation at all resolved wavelengths.
Time advancement is performed by means of a standard three-stage third-order
explicit Runge-Kutta algorithm.

\subsection{Computational domain}

\begin{figure}
    \centering{\includegraphics[width=13cm]{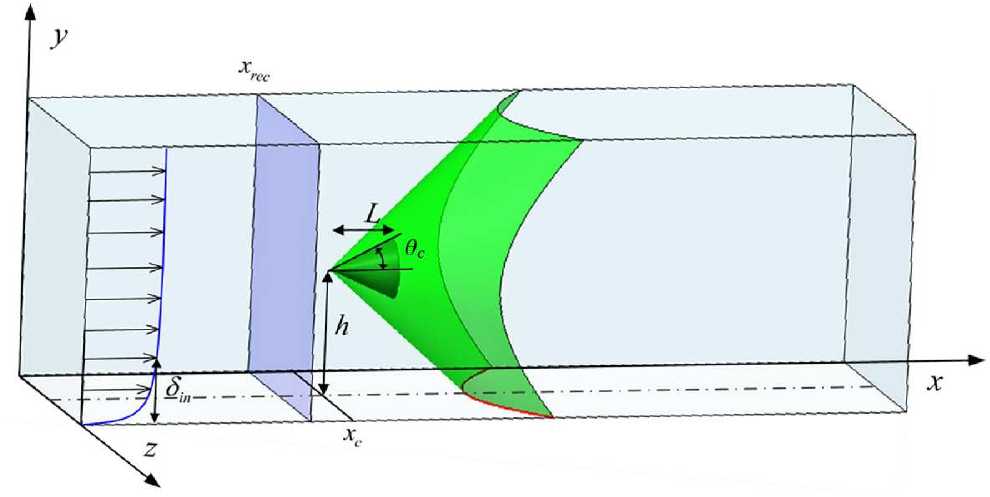}}
	\caption{Sketch of computational domain for CSBLI analysis.
	$\delta_{in}$ is the inflow boundary layer thickness, 
        $x_{rec}$ is the boundary layer recycling station, $ {x_c} $ is the $x$ coordinate of the
	cone leading edge. The green surface depicts the conical shock,
        whose wall trace is highlighted in red.}  
	\label{fig:domain}
\end{figure}

The computational domain employed for the simulation is sketched in
figure~\ref{fig:domain}. 
The choice of the streamwise extent (${L_x}=150{\delta _{in}}$, with $\delta_{in}$ the inflow
boundary layer thickness)
is dictated by the requirement of having a sufficient boundary layer development region 
past the inflow and to analyze the recovery region past the interacting shock.
The spanwise length (${L_z}=60{\delta _{in}}$) guarantees
that no spurious coherence develops in the upstream boundary layer, and allows 
to minimize numerical blocking from the side boundaries.
The domain height (${L_y}=30{\delta _{in}}$) is such that numerical
wave reflections from the upper boundary are also minimized.
Based on previous experience with DNS of SBLI and of a set of preliminary simulations,
the domain has been discretized with a grid including
$1536 \times 384 \times 1280$ nodes. Uniform spacing is used in the
streamwise and spanwise directions, whereas nodes are clustered in the wall-normal ($y$)
direction according to a hyperbolic sine stretching function up to $y/\delta_{in} = 6.5$.
In terms of wall units evaluated in the undisturbed boundary layer upstream of the interaction zone,
the streamwise and spanwise spacings are $\Delta x^+ \approx 10$, 
$\Delta z^+ \approx 5$, respectively, whereas the spacing in the
wall-normal direction ranges between $0.7$ at the wall and $12$.
Here and elsewhere, the + subscript is used to denote normalization 
with respect to the friction velocity $u_\tau = \sqrt{\tau_w/\rho_w}$
(where $\tau_w$ and $\rho_w$ are the wall shear stress and density),
and the viscous length scale $\delta_v = \nu_w / u_\tau$ 
(where $\nu_w$ is the wall kinematic viscosity).
A further a-posteriori check has shown that the grid spacing
in each coordinate direction is nowhere larger than about 
three local Kolmogorov length scales.

\subsection{Treatment of shock generator}

\begin{figure}
  \centering
  \psfrag{a}[l][][0.8]{{\color{green} R}}
  \psfrag{b}[l][][0.8]{{\color{red} I}}
  \psfrag{c}[l][][0.8]{{\color{blue} W}}
  \includegraphics[width=7.cm,clip]{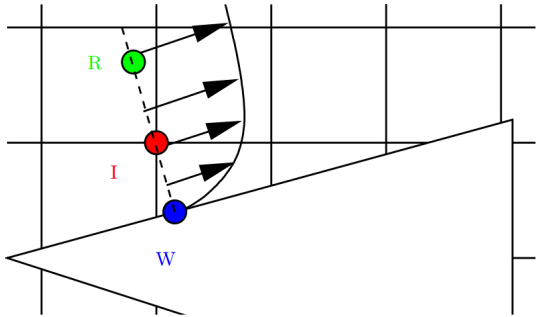}
  \vskip1.em
  \caption{Sketch of the immersed-boundary treatment. I is the interface node, and W and R are the corresponding wall-normal foot and reflected control point.}
  \label{fig:IB}
\end{figure}

In order to accommodate the conical shock generator in the Cartesian mesh used for the
DNS we rely on the immersed boundary method, using the 
direct forcing approach of~\citet{Fadlun2000-Combined}, in
which the Navier-Stokes equations are replaced by suitable interpolation 
formulas at the interface nodes between solid and fluid.
A geometrical preprocessor based on the ray tracing algorithm~\citep{orourke98} is applied to 
tag solid and fluid nodes.
Once the fluid nodes are located, we identify the subset of nodes for which discretization of the Navier-Stokes equations
involves the use of solid nodes, which are tagged as interface nodes and used to indirectly prescribe
desired values to the conservative variables conditions at the fluid/body interface.
For that purpose (see figure~\ref{fig:IB}), for each interface node (I)
a wall-normal line is considered, along which a control point (R) is considered, at fixed distance
from the wall foot (W), where the flow variables are interpolated from neighboring grid nodes.
An equilibrium wall function is then used to define the flow variables at I based on those at R~\citep{Tessicini2002-wall}.
Implementation details are provided in \citet{Bernardini2016-On}.
The cone is resolved with approximately 72, 56 and 138 grid intervals in the
streamwise, wall-normal and spanwise directions, respectively, and
the distance of the reflected control points from the nearest wall expressed in local wall units along the cone ranges between 3 and 70.

\subsection{Flow conditions}

The flow conditions are selected to be as close as possible to the
reference experiment of~\citet{Hale2015-Interaction}. 
The shock is generated by a circular cone with height $L/\delta_{in}=6.835$ 
and half opening angle is $\theta_c = 25^\circ$.
The apex of the cone is at ${x_{c}} = 30{\delta _{in}}$,
and its axis is parallel to the wall at a distance $h/ \delta_{in}=13.67$ (see figure~\ref{fig:domain}).
The upstream flow has Mach number $M_{\infty} = 2.05$, and the Reynolds number based on
the inflow boundary layer thickness is $\Rey_{\delta_{in}}=5000$. The latter is about 
a factor of fifty less than the reference experiment, but as will be shown later 
this is not the cause of major quantitative differences.

A critical issue in the numerical simulations of spatially developing flows 
is the prescription of suitable inflow conditions to achieve a fully developed
boundary layer state within the shortest possible fetch. In the present DNS this is achieved
through a rescaling-recycling procedure~\citep{Xu2004-Assessment},
whereby a cross-stream slice of the flow field is extracted at every Runge-Kutta sub-step
at the recycling station $x_{rec}$, and fed back to the inflow upon suitable rescaling.
This approach generates a realistic turbulent boundary layer within a
short distance from the inflow, and it allows to control skin friction and
thickness of the simulated boundary layer.
To minimize spurious time periodicity that may result from application of
quasi-periodic boundary conditions in the streamwise direction, the
recycling station is set at $x_{rec} = 25 \delta_{in}$, 
also sufficiently upstream of the cone apex.
Non-reflecting characteristic boundary conditions
are applied to the outflow, at the top boundary, and in the spanwise direction for $x>x_{rec}$,
whereas spanwise periodicity is assumed for $x \le x_{rec}$.
Unsteady characteristic boundary conditions are specified at the bottom no-slip wall~\citep{Poinsot1992-Boundary},
with temperature set to its adiabatic value. 
Flow statistics have been collected from time
$ t_0 u_\infty / \delta _{in} \approx 216 $
to time $ t_f u_\infty / \delta _{in} \approx 1572$,
at intervals of $\Delta t u_\infty / \delta _{in} \approx 0.1$.
The long sampling time is needed to achieve convergence 
of point-wise statistics in time in the absence of homogeneous directions,
and it makes the present calculation quite time-consuming.
The statistical analysis is carried out by splitting the instantaneous
quantities into their mean and fluctuating components,
using either the standard Reynolds decomposition ($ f=\overline f  + f' $)
or the density-weighted (Favre) decomposition ($ f = \widetilde f + f'' $),
where $ \widetilde f = \overline {\rho f} /\overline \rho $.

\section{Results}\label{sec:results_and_discussion}

\subsection{Characterization of the incoming boundary layer}

\begin{table}
 \setlength\tabcolsep{2.5mm}
  \begin{center}
    \def~{\hphantom{0}}
    \begin{tabular}{lcccccccc}
      $M_e$  & $\Rey_\theta$ & $\Rey_{\delta_2}$ & $\Rey_\tau$ & $C_f$                 & $\delta^*/\delta$ & $\theta/\delta$ & $H$    & $H_i$  \\[3pt]
      $2.05$ & $630$         & $410$             & $160$       & $3.63 \times 10^{-3}$ & $0.326$           & $0.082$         & $3.97$ & $1.60$ \\
    \hline
    \end{tabular}
    \caption{Properties of incoming turbulent boundary layer at the
    reference station $x_{ref} = 27.5 \delta_{in}$.
    $M_e = u_e/c_e$; $\Rey_\theta = \rho_e u_e \theta/\mu_e$;
    $\Rey_{\delta_2} = \rho_e u_e \theta/\mu_w$;
    $\Rey_\tau = \rho_w u_\tau \delta / \mu_w$;
    $C_f = 2 \tau_w /(\rho_e u_e^2)$;
    $H = \delta^* / \theta$;
    $H_i = \delta_i^* / \theta_i$. 
    The subscript $e$ denotes properties evaluated at the edge of the boundary layer.}
    \label{tab:blprop}
  \end{center}
\end{table}

A necessary preliminary check is that the incoming boundary layer is properly developed
prior to interaction with the conical shock.
For that purpose, we consider a reference station at $x_{ref} = 27.5 \delta_{in}$,
located upstream of the cone leading edge.
The global boundary-layer properties at this station are listed in table~\ref{tab:blprop}, where
$\delta$ is the $99\%$ thickness,
$\delta^*$ is the displacement thickness,
\begin{equation}
\delta^* = \int_0^{{\delta_e}} \left( 1 - {{\bar \rho } \over {{\rho_e}}}{{\bar u} \over {{u_e}}} \right) \diff y ,
\label{eqn:dstar}
\end{equation}
and $\theta$ is the momentum thickness,
\begin{equation}
\theta  = \int_0^{{\delta_e}} {\frac{{\bar \rho }}{{{\rho_e}}}\frac{{\bar u}}{{{u_e}}}\left( {1 - \frac{{\bar u}}{{{u_e}}}} \right)} \diff y ,	
\label{eqn:theta}
\end{equation}
the upper integration limit $\delta_e$ denoting the edge of the rotational part of the boundary layer,
defined as the point where the mean spanwise vorticity becomes less than $0.005{u_\infty }/{\delta _{in}}$~\citep{Pirozzoli2010-Direct}.
The subscript $e$ is used to denote the corresponding external flow properties.
The table also reports equivalent incompressible boundary layer properties evaluated 
by setting to unity the density ratios in equations~\eqref{eqn:dstar}, \eqref{eqn:theta}, 
and referred to with the subscript $i$.

\begin{figure}
  \centering
  \psfrag{x}[][][1.0]{$ y^+ $}
  \psfrag{y}[][][1.0]{$ u_{vd}^ + $}
  \psfrag{u}[][][1.0]{$\tau_{ij}^*$}
  (a)\includegraphics[width=6.2cm,angle=0,clip]{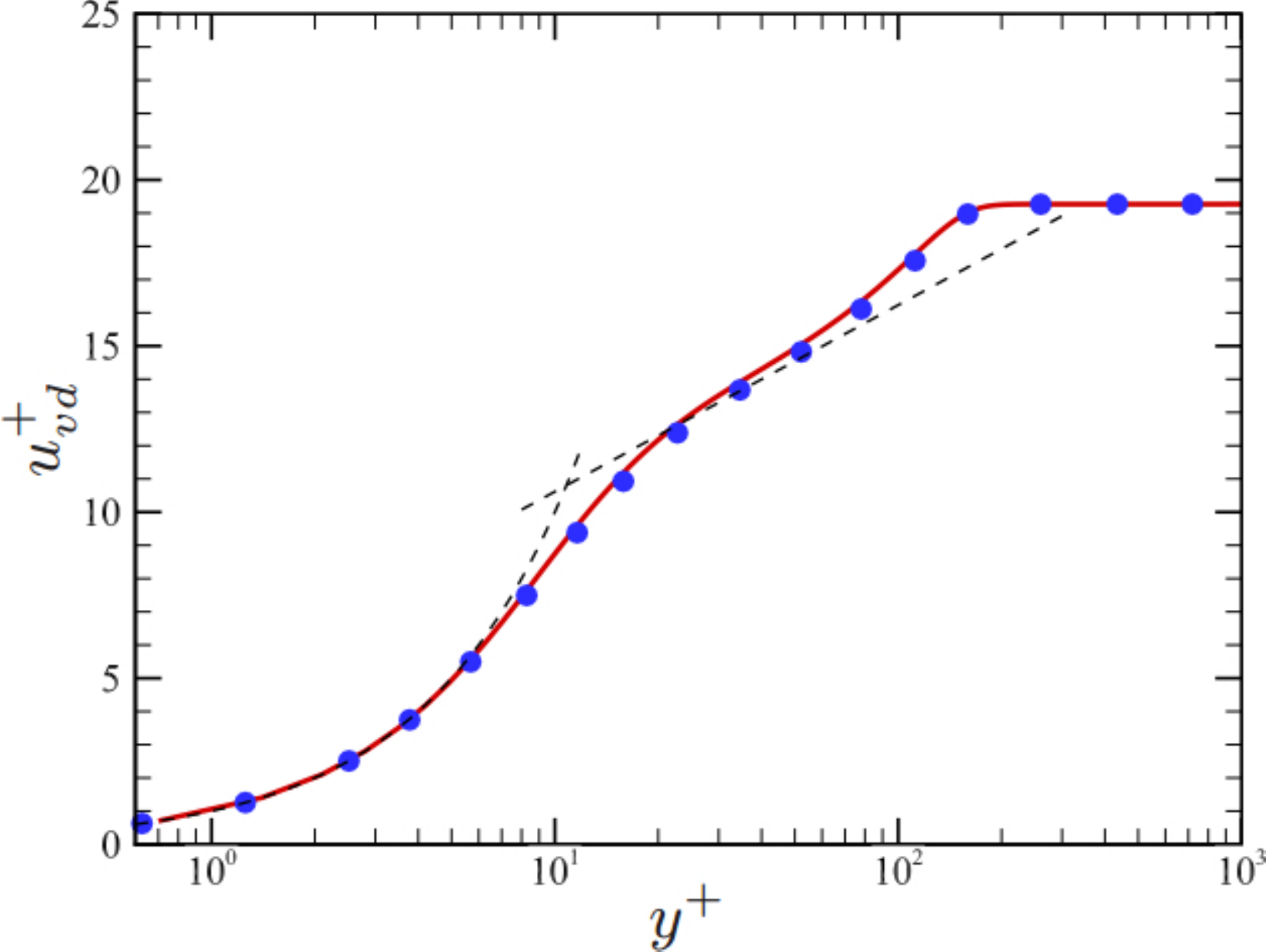}
  (b)\includegraphics[width=6.2cm,angle=0,clip]{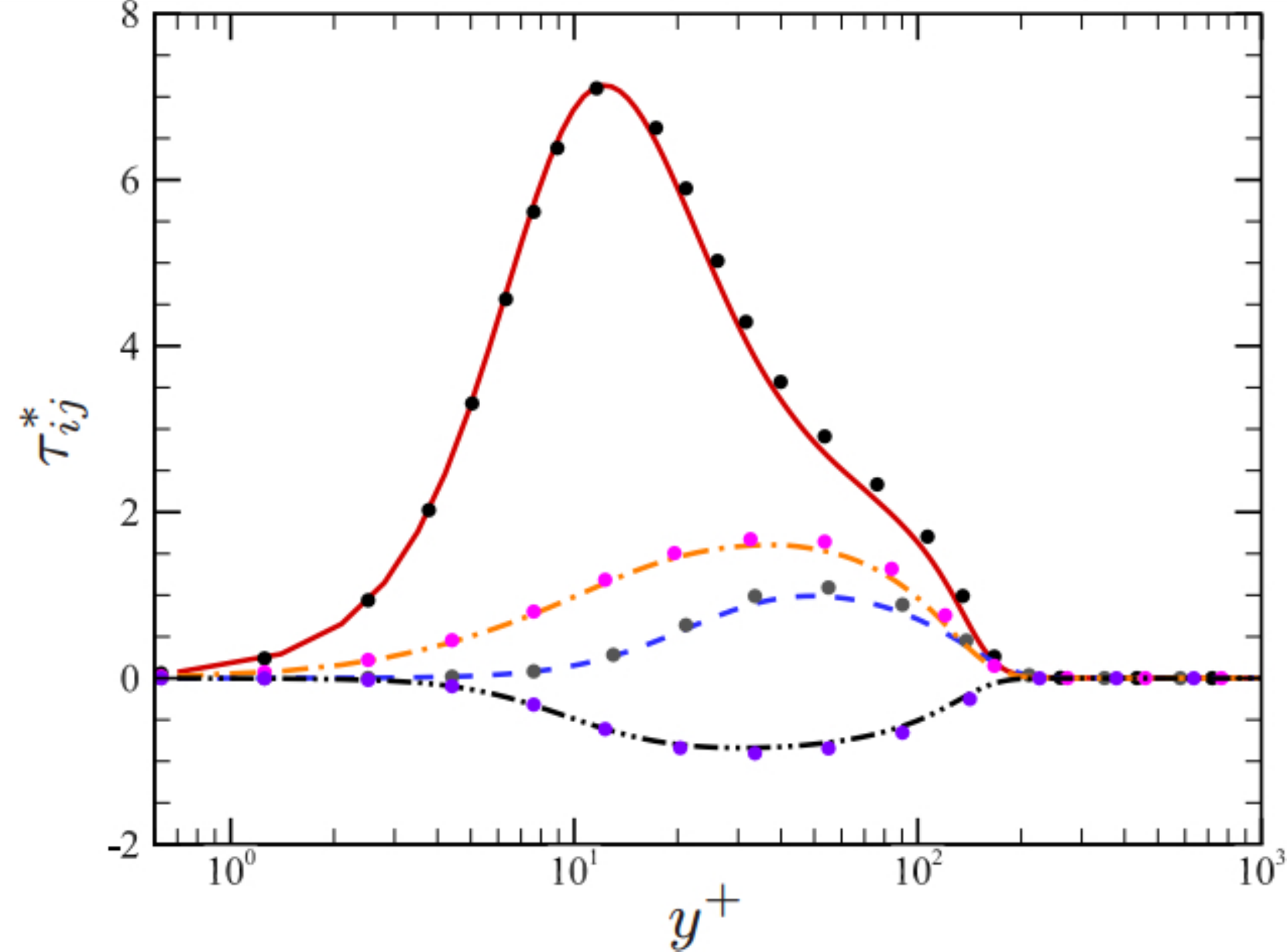} \vskip1.em
  \caption{Van Driest transformed mean streamwise velocity (a),
  and density-scaled turbulent stresses (b) at the reference station
  ($ {x_{ref}} = 27.5{\delta _{in}} $). Lines refer to the present DNS data,
  and symbols to reference data~\citep{Pirozzoli2011-Turbulence}.
  In panel (a) the dashed line denotes a compound of
  $u^+ = y^+$ and $u^+ = 5.0 + 1/0.41 \log y^+$.
  In panel (b) we show $\tau_{11}^*$ (solid), $\tau_{22}^*$ (dashed); $\tau_{33}^*$ (dash-dot); $\tau_{12}^*$ (dash-dot-dot).}
 \label{fig:VD}
\end{figure}

The van Driest effective velocity, 
\begin{equation}
u_{vd} = \int_0^{\bar u}{\left({\bar\rho} \over \bar\rho_w\right)}^{1/2}d\bar u,
\label{equation11}
\end{equation}
is used in figure~\ref{fig:VD}a to compare with boundary layer 
data at similar flow conditions~\citep{Pirozzoli2011-Turbulence}.
Comparison with the reference data is quite good, and
in particular the velocity profiles show linear behavior up to $y^+ \simeq 5$,
as expected for adiabatic boundary layers~\citep{Smits-2006-Turbulent},
and a narrow range with near logarithmic variation.
Comparison of the density-scaled velocity correlations
\begin{equation}
  \tau_{ij}^* = \frac{ \overline{\rho} \widetilde{u''_i u''_j}}{\tau_w} ,
  \label{density_scaled}
\end{equation}
is also shown in figure~\ref{fig:VD}b.
The agreement with the reference data is again quite good, which leads us to
conclude that the upstream boundary layer well corresponds to a healthy state
of equilibrium wall turbulence.

\subsection{General flow organization}

\begin{figure}
	\centering
	(a) \includegraphics[width=10cm,clip]{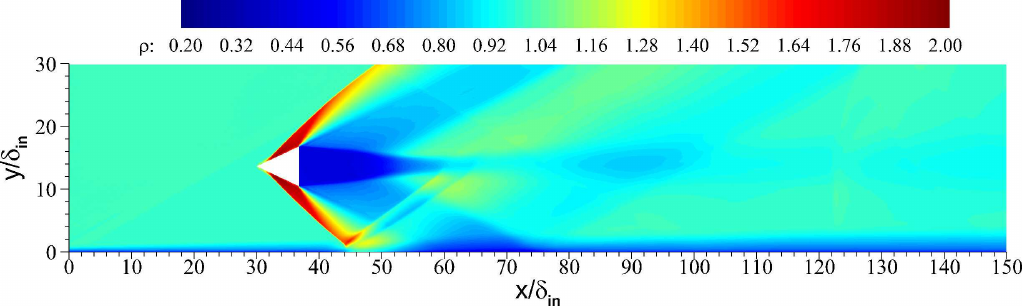}
	\\
	(b) \includegraphics[width=10cm,clip]{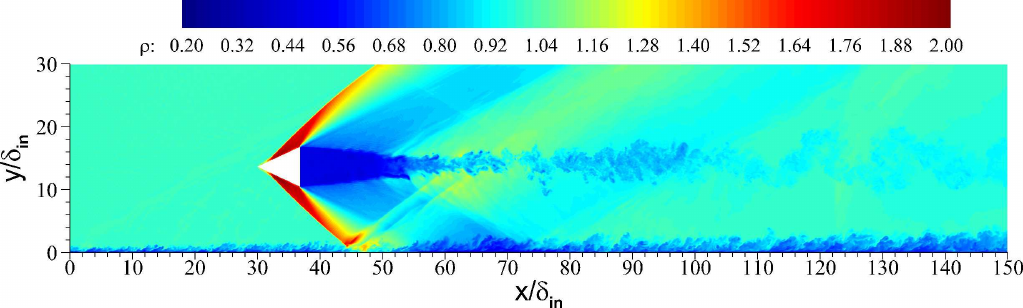}
	\\	
	(c) \includegraphics[width=10cm,clip]{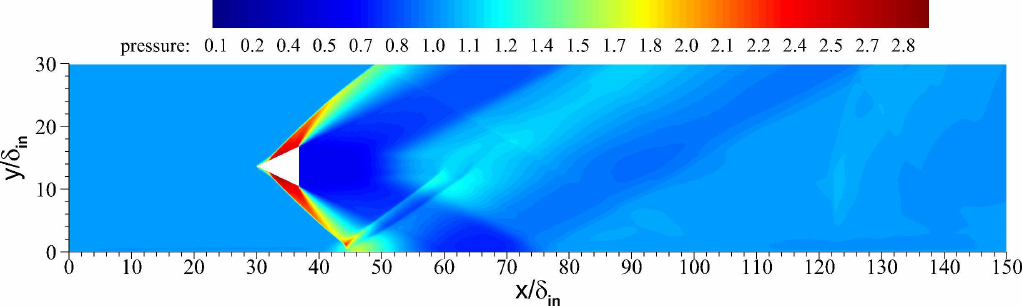}
	\\
	(d) \includegraphics[width=10cm,clip]{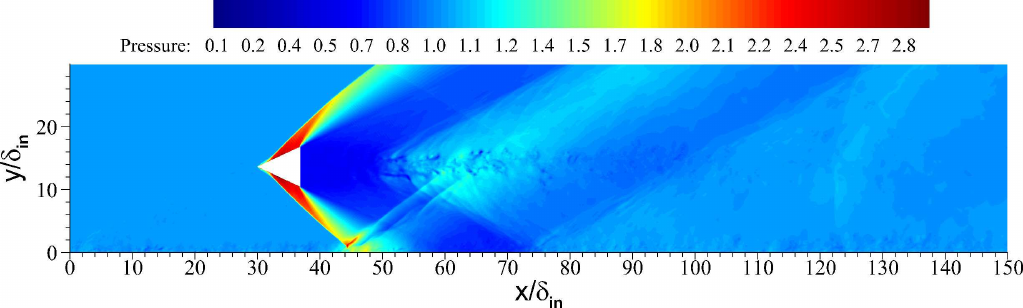}
	\\
	(e) \includegraphics[width=10cm,clip]{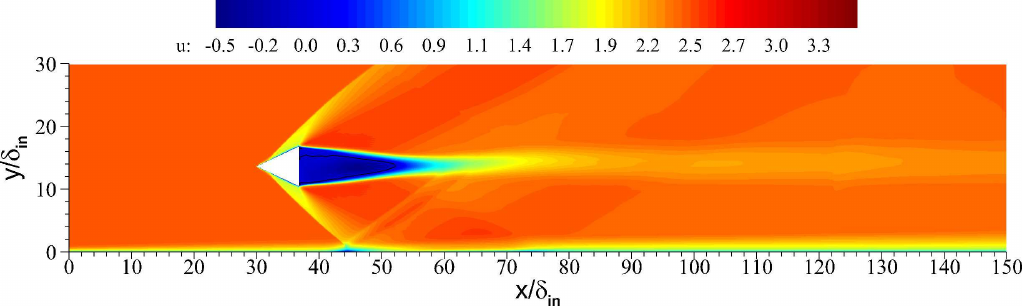}
	\\
	(f) \includegraphics[width=10cm,clip]{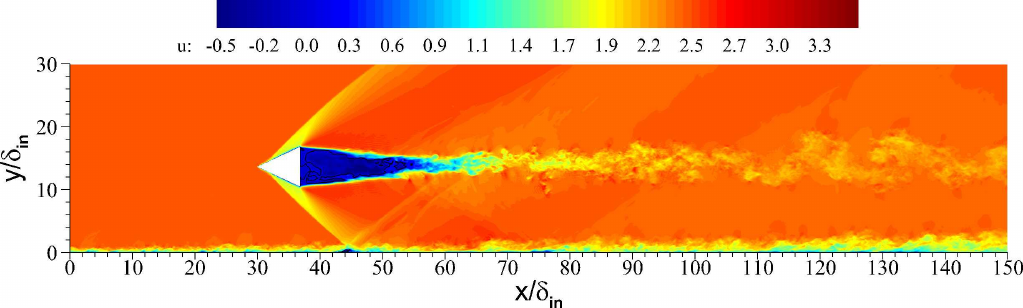}
        \vskip1.em
	\caption{Visualization of time-averaged (a, c, e) and instantaneous (b, d, f) flow field
	  in the symmetry plane ($z=0$). The blank region corresponds to the conical shock generator.
	  Panels (a, b): density, $ 0.2 \le \bar \rho /{\rho _\infty } \le 2.0 $;
	  panels (c, d): pressure, $ 0.1 \le \bar p/{p_\infty } \le 2.9 $; 
	  panels (e, f): streamwise velocity, $- 0.5 \le \bar u/{u_\infty } \le 3.5 $. 
	  Color scale from blue to red.}
	\label{fig:mean_symmetry}
\end{figure}

Having established the properties of the incoming boundary layer, we proceed to describe the main 
features of the flow upon interaction with the conical shock.
For that purpose time-averaged and instantaneous fields of density, pressure and streamwise velocity
are shown in the symmetry plane in figure~\ref{fig:mean_symmetry}. 
The flow exhibits an overall organization similar to experimental observations~\citep{Hale2015-Interaction}, 
with a shock wave generated by cone which
impinges on the bottom wall surface, thus leading to CSBLI. 
This is clearly seen in density field, which 
well highlights the overall wave pattern. 
The main interacting shock is generated as straight at the cone surface,
and subsequently it bends backward upon merging with the 
expansion fan which arises at the cone trailing edge.
A continuous compression may be inferred 
from the pressure contours along the cone surface, which may be justified recalling that 
the flow is not uniform past conical shocks, unlike for planar shocks.
The shock impinges on the bottom wall 
at $x/{\delta _{in}} \approx 45$.
The streamwise velocity contours in figure~\ref{fig:mean_symmetry}e well
highlights the thickening of the boundary layer in this region.
As also in canonical planar interactions~\citep{Pirozzoli2006-Direct}, the reflected wave system
consists of a precursor wave associated with the upstream influence mechanism,
and a trailing wave associated with boundary layer reattachment.
The figures also bring out the presence 
of an extended low-speed region in the wake of the cone with turbulent flow,
which is closed by a conical recompression shock~\citep{Herrin1994-Supersonic}, originating
at $x / \delta_{in} \approx 50$, and interacting with the developing boundary layer 
at $x / \delta_{in} \approx 75$.
An additional weak compression wave is also seen to be generated at the upper
boundary of the computational domain owing to imperfect radiation of numerical
disturbances, and hitting the wall at $x / \delta_{in} \approx 100$.
The dynamical character of the reflected shock system, which oscillates 
back and forth may be clearly appreciated in an attached supplementary movie.

\begin{figure}
	\centering
        (a)
	\includegraphics[width=10cm]{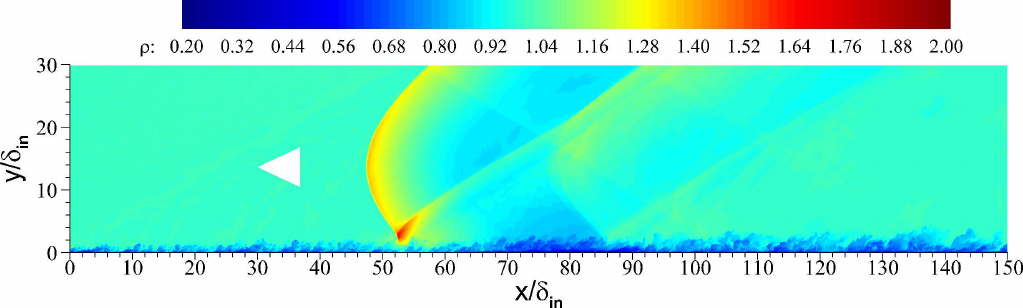}
	\\
        (b)
	\includegraphics[width=10cm]{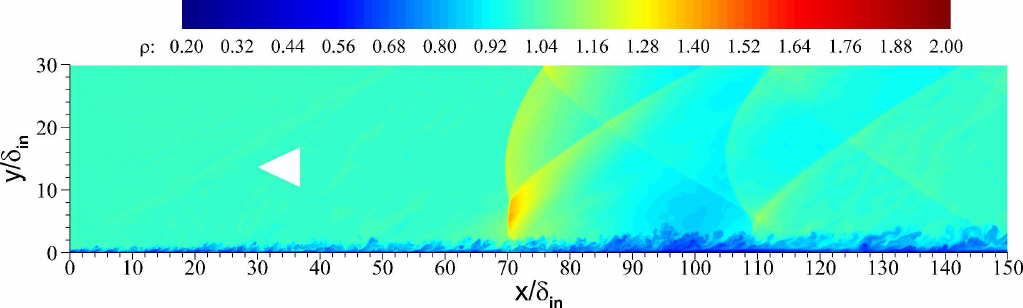}
        \vskip1.em
	\caption{Instantaneous density contours at 
        $ z/{\delta _{in}} = -15 $ (a), $ z/{\delta _{in}} = -30 $ (b).
	Contours are shown for $ 0.2 \le \rho /{\rho _\infty } \le 2.0 $, from blue to red.
	The cone trace in the symmetry plane is shown for reference purposes.}
	\label{fig:rho_xyplanes}
\end{figure}

To understand the three-dimensional character of the flow field, 
wall-normal/streamwise sections at two additional spanwise locations 
are reported in figure~\ref{fig:rho_xyplanes},
where instantaneous density contours are plotted using the same style as in figure~\ref{fig:mean_symmetry}. 
The figure shows that the conical shock traces are not straight lines as in the symmetry plane,
but rather they have hyperbolic shape, as should be the case.
Additional features should be also noted.
In particular, we find that reflection of the primary interaction shock 
changes from regular to Mach type~\citep{Migotsky1951-Three-Dimensional}. 
Second, the density jump across the shock decreases gradually owing to flow three-dimensionality
and interaction of the expansion fan originating at the cone trailing edge.
Third, the wake recompression shock has a similar structure as the primary conical
shock, and its in-wall trace is also hyperbolic.

\begin{figure}
	\centering
    (a)\includegraphics[width=3.3cm]{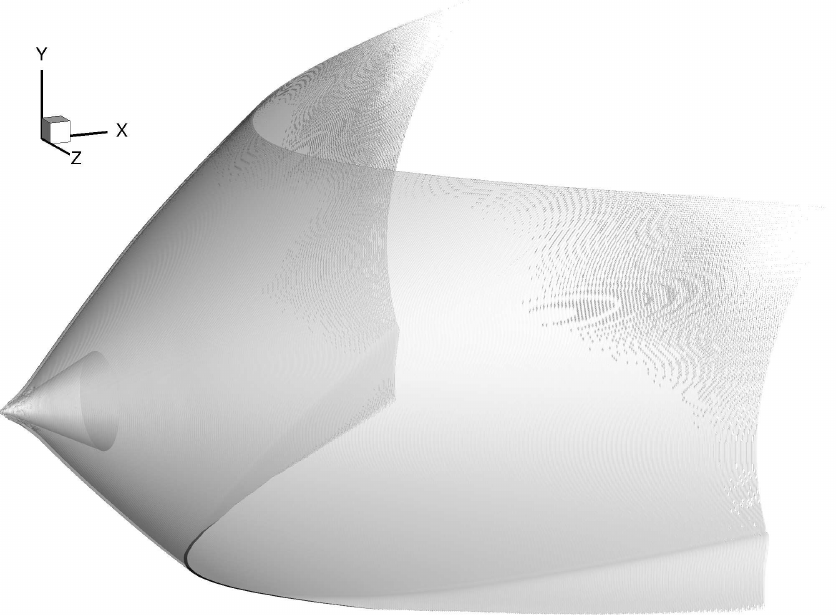}
    (b)\includegraphics[width=5.4cm]{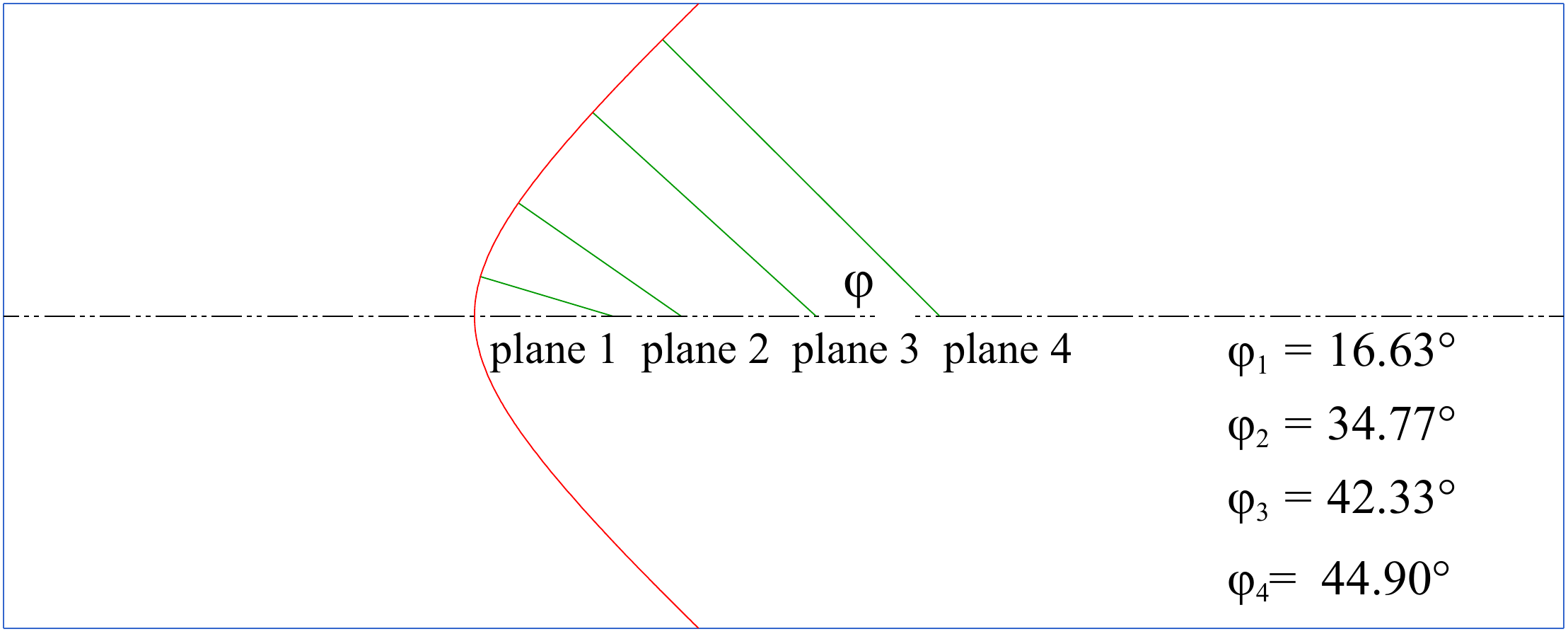}
    (c)\includegraphics[width=3.3cm]{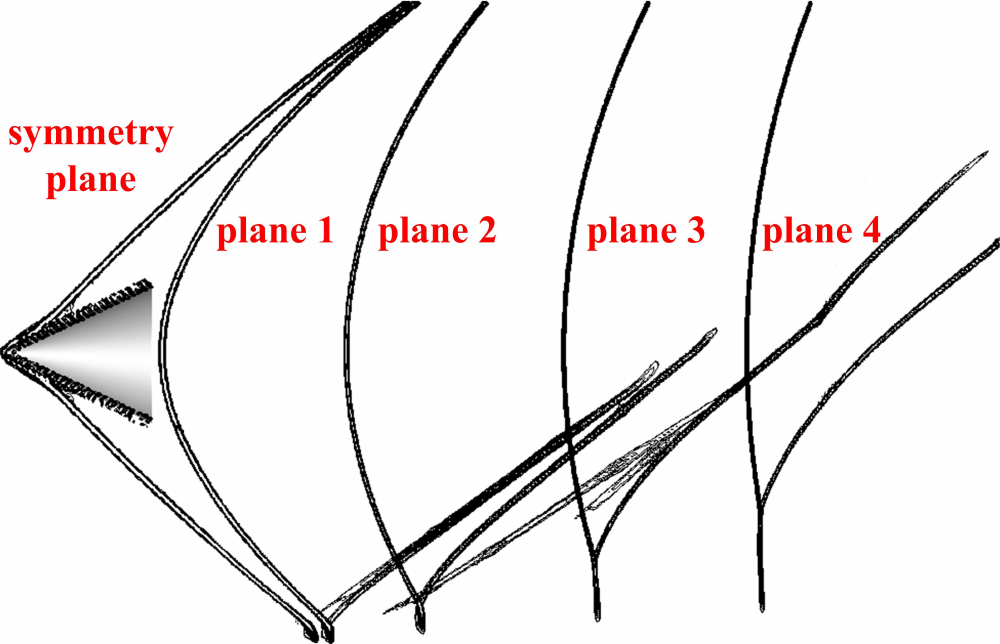}
	\\ \vskip2.em
    (d)\includegraphics[width=4.0cm]{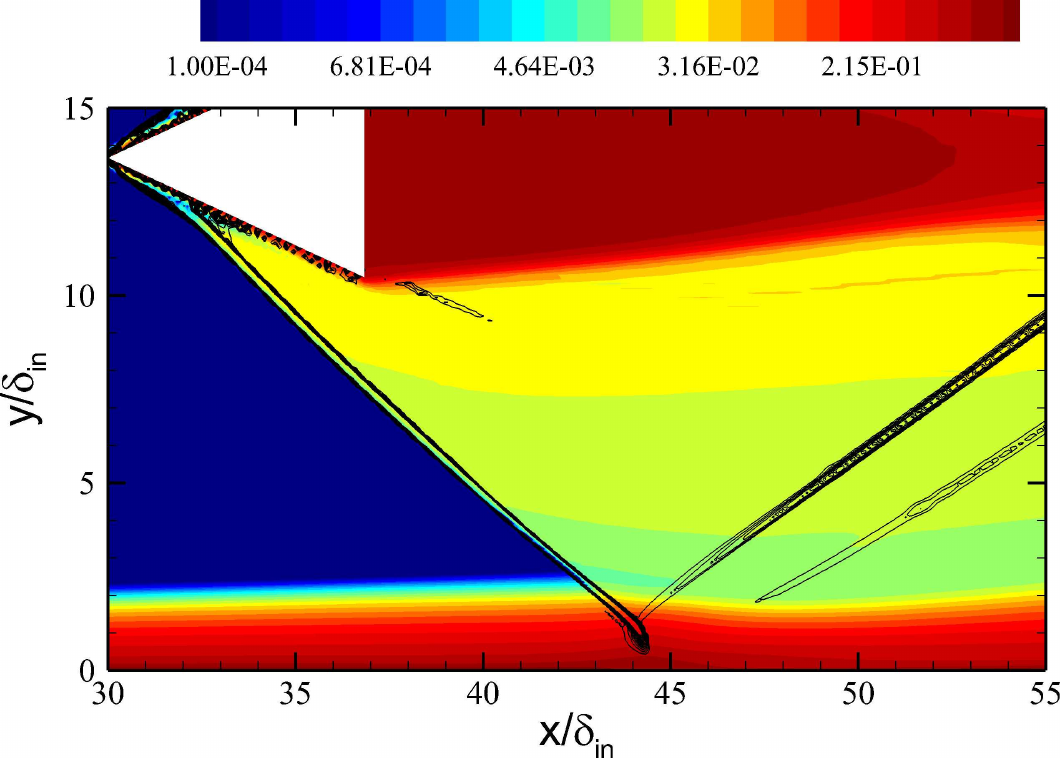}
    (e)\includegraphics[width=4.0cm]{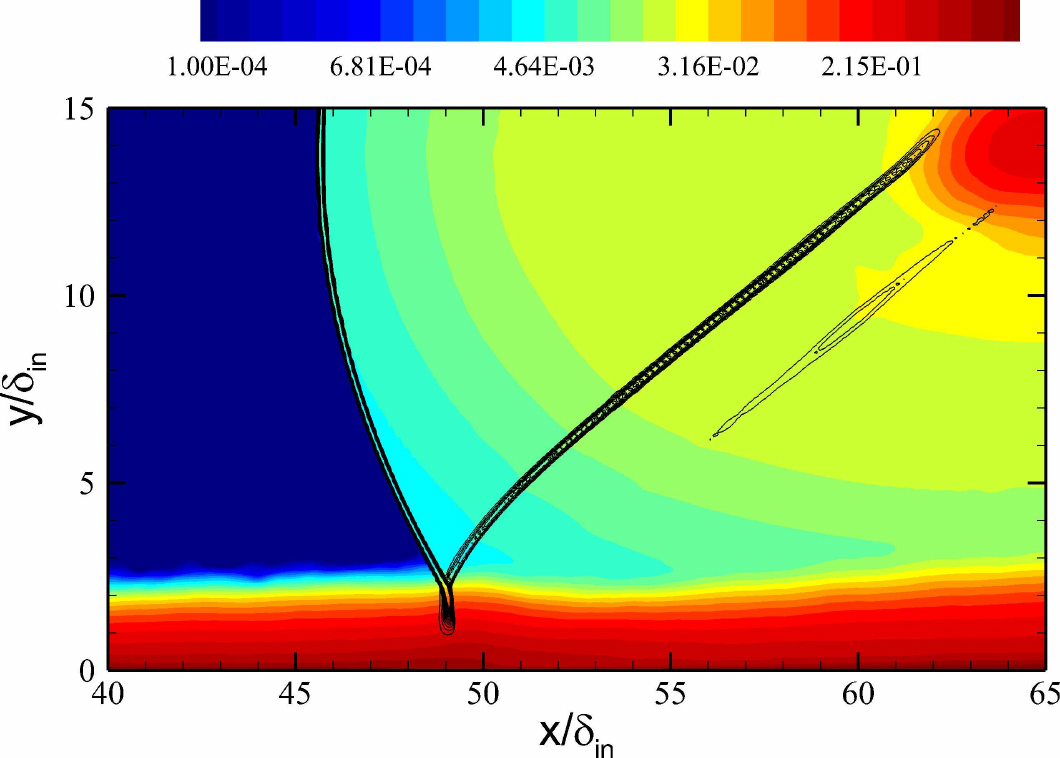}
    (f)\includegraphics[width=4.0cm]{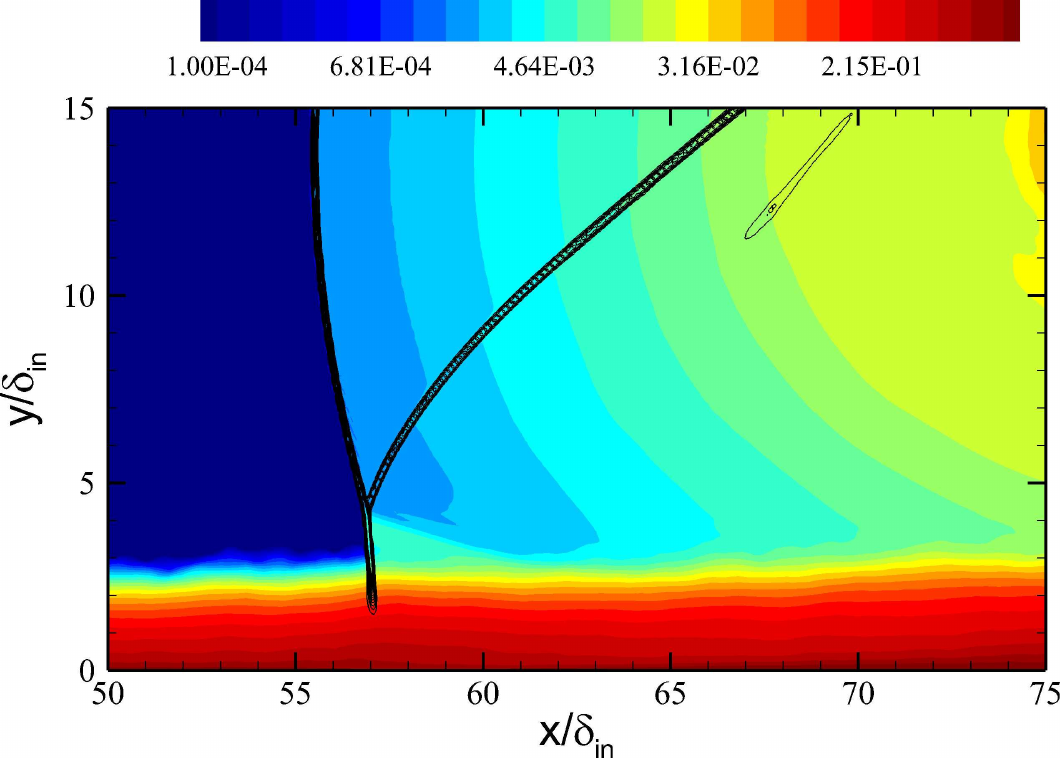}
	\\ \vskip1.em
	\caption{Analysis of shock structure:
   	 iso-surface of shock sensor (a);
	 nomenclature for analysis in wall-normal planes (b);
         shock sensor in control planes (c);
	 entropy contours in:
	 symmetry plane (d), 
	 plane 2 (e),
	 plane 3 (f).
         Entropy levels are logarithmically spaced in the range $10^{-5} \le s \le 1.0$.}
	\label{fig:Mach_refl}
\end{figure}

\citet{Migotsky1951-Three-Dimensional} theoretically analyzed the reflection of a conical shock
from a planar surface in the limit of strictly inviscid flow.
They predicted that transition from regular to Mach reflection should occur at a critical value of 
the angle formed by normal to the shock trace in the wall plane with the streamwise direction 
($\varphi$, see figure~\ref{fig:Mach_refl}b). 
At the flow conditions under scrutiny here, transition should occur for $\varphi \approx 9^{\circ}$.
In order to visualize the shock structure in the present DNS,
in figure~\ref{fig:Mach_refl} we show contours of the shock sensor.
Information is also provided in figure~\ref{fig:Mach_refl} in a series of planes normal to the 
shock wall trace.
To clarify the occurrence of different types of shock reflection, in panels
(d-f) we show contours of the thermodynamic entropy ($s = p/\rho^{\gamma}$).
The analysis is here made complicated by the presence of the boundary layer,
which is itself the cause for substantial increase of entropy from the free stream
toward the wall, and which causes refraction of the shock wave as it penetrates
layers with lower speed.
Based on the available data, we may conclude that reflection is likely to be of regular
type for control planes 1 and 2, whereas it is certainly of Mach type starting at 
control plane 3, for which a distinct triple point emerges outside the boundary layer.
Accordingly, higher entropy is observed for fluid particles traveling underneath the triple point 
than above.

\begin{figure}
	\centering
	\includegraphics[width=13.0cm]{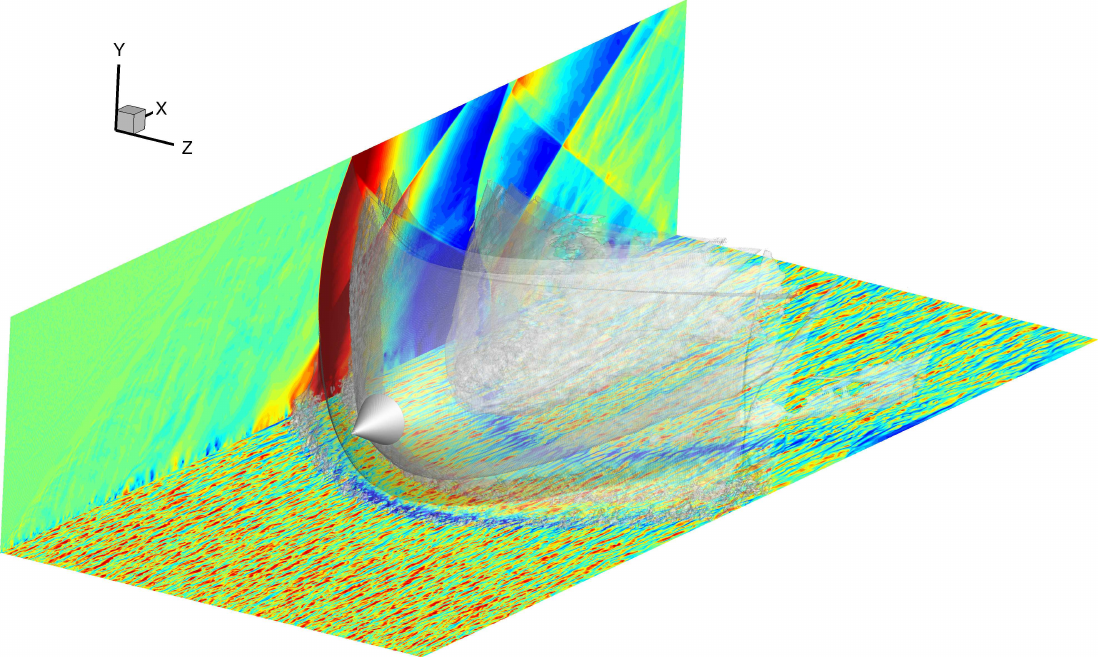}
	\caption{Three-dimensional view of CSBLI. The shock structure is educed through the pressure iso-surface $p=1.1 {p_\infty }$. Streamwise velocity contours are shown for $ -0.3 < u/{u_\infty } < 2.3 $ (color scale from blues to red) in a near-wall plane at $ y^+ = 10.5 $. Pressure contours are shown in a side plane for $ 0.8 < p/{p_\infty } < 1.2 $ (color scale from blue to red). Supplementary movie 2 is available for this figure.}
	\label{fig:shockshape}
\end{figure}

A three-dimensional rendering of the flow structures is provided in 
figure~\ref{fig:shockshape}, which includes iso-surfaces of pressure,
streamwise velocity contours in a near-wall plane, and pressure contours in a side plane. 
This figure qualitatively brings out the strong relationship between the shock system and the boundary layer evolution, 
also observed in planar shock interactions~\citep{Aubard2013-Large-Eddy,Pirozzoli2011-Direct}.
Specifically, figure~\ref{fig:shockshape} shows the presence of elongated streaks of high- and low-speed momentum 
in the ZPG region. As the boundary layer penetrates the first APG region it experiences strong retardation, and a region
of low momentum forms past the upstream branch of the interacting shock.
Streaks are found to reform quickly past the reflected shock, 
and they undergo a second suppression/reformation cycle across the second APG zone 
associated with wall impingement of the recompression shock.

\begin{figure}
	\centering
	\includegraphics[width=14.0cm]{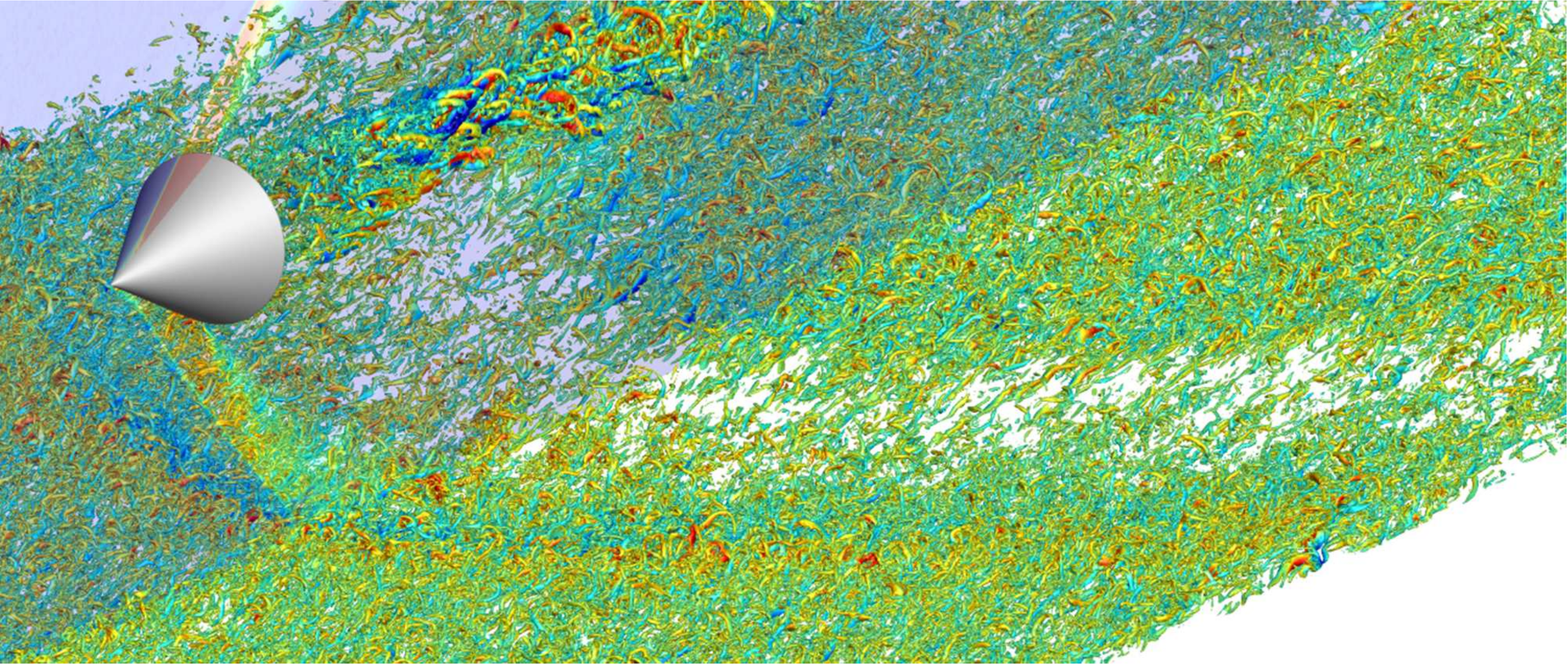}
	\caption{Vortical structures in CSBLI. Vortices are educed through the iso-surface of the swirling strength ($\lambda _{ci} = 1.24 u_{\infty}/\delta_{in}$), and colored with the wall-normal velocity ($ -0.37 < v/u_{\infty} < 0.37 $, color scale from blue to red). Supplementary movie 3 is available for this figure.}
	\label{fig:3d_vortex}
\end{figure}

Interaction of the boundary layer with a conical shock has a strong impact on the 
structure and number of vortical eddies populating the near-wall region.
In figure~\ref{fig:3d_vortex} we report a three-dimensional rendering of vortical structures
detected as iso-surfaces of the swirling strength, defined as the imaginary part of the
complex conjugate eigenvalue of the velocity gradient tensor~\citep{Zhou1999-Mechanisms}.
Visual inspection of many flow samples (supplementary movie 3) shows a vortex population
in the ZPG region which is similar to what found in incompressible boundary layers~\citep{Wu-2009-Direct},
including some hairpin-shaped vortices as well as more asymmetric, cane-shaped vortices.
The vortices tend to disappear in the FPG region, and to reform past the recompression shock.
Many hairpin-shaped vortices are also observed in the wake region past the shock generator,
which propagate downstream evolving into ring-shaped eddies.

\subsection{Wall pressure}

\begin{figure}
	\centering
	\includegraphics[width=10.0cm]{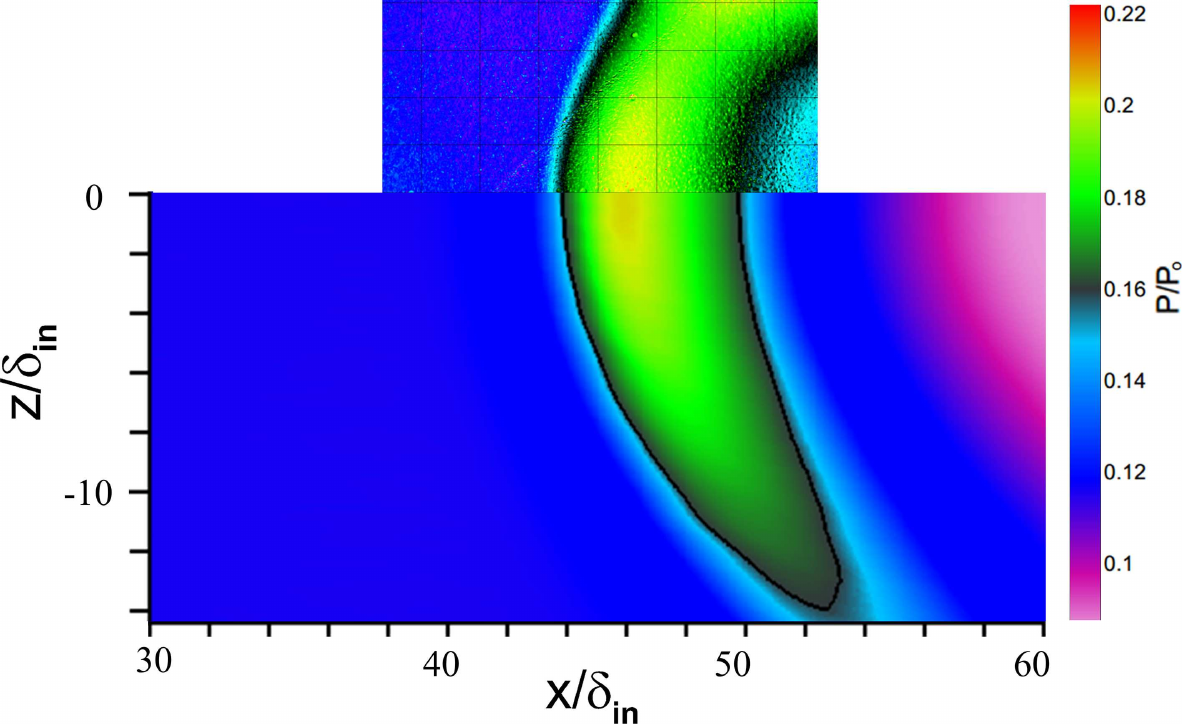} 
	\caption{Mean wall pressure as obtained from experiments~\citep[upper half]{Hale2015-Interaction}, and
	  from DNS (lower half). Contour levels are shown between $ 0.08<p/p_0<0.22 $, with $p_0$ the free-stream total pressure, the black line denoting the $ p = 0.16 p_{0}$ iso-line.}
	\label{fig:PSP}
\end{figure}

\begin{figure}
	\centering
	\psfrag{x}[][][1]{$ x/\delta_{in} $}
	\psfrag{p}[][][1]{$ p/{p_\infty } $}
        (a)
	\includegraphics[width=6.0cm]{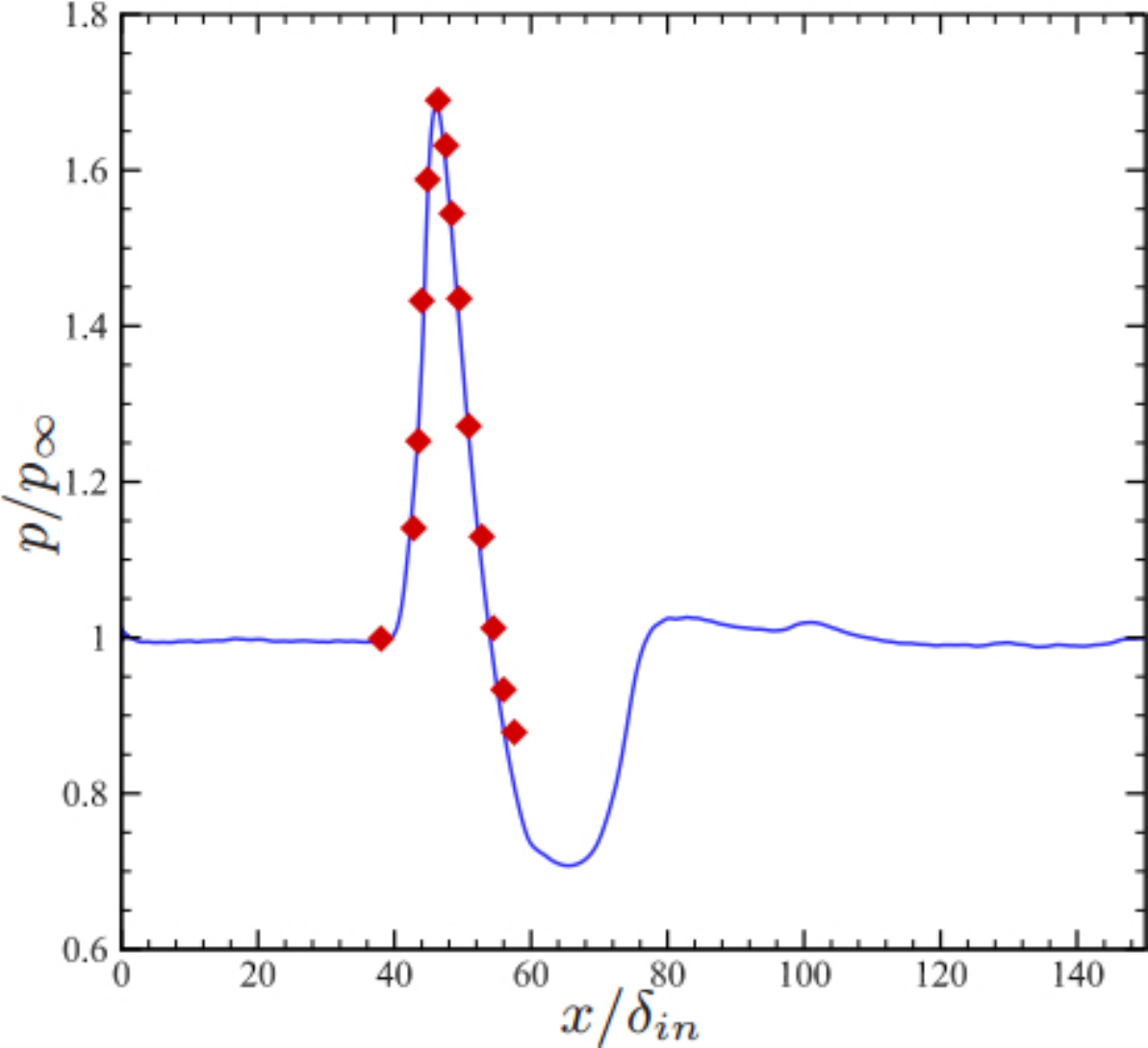}
        (b)
	\includegraphics[width=6.0cm]{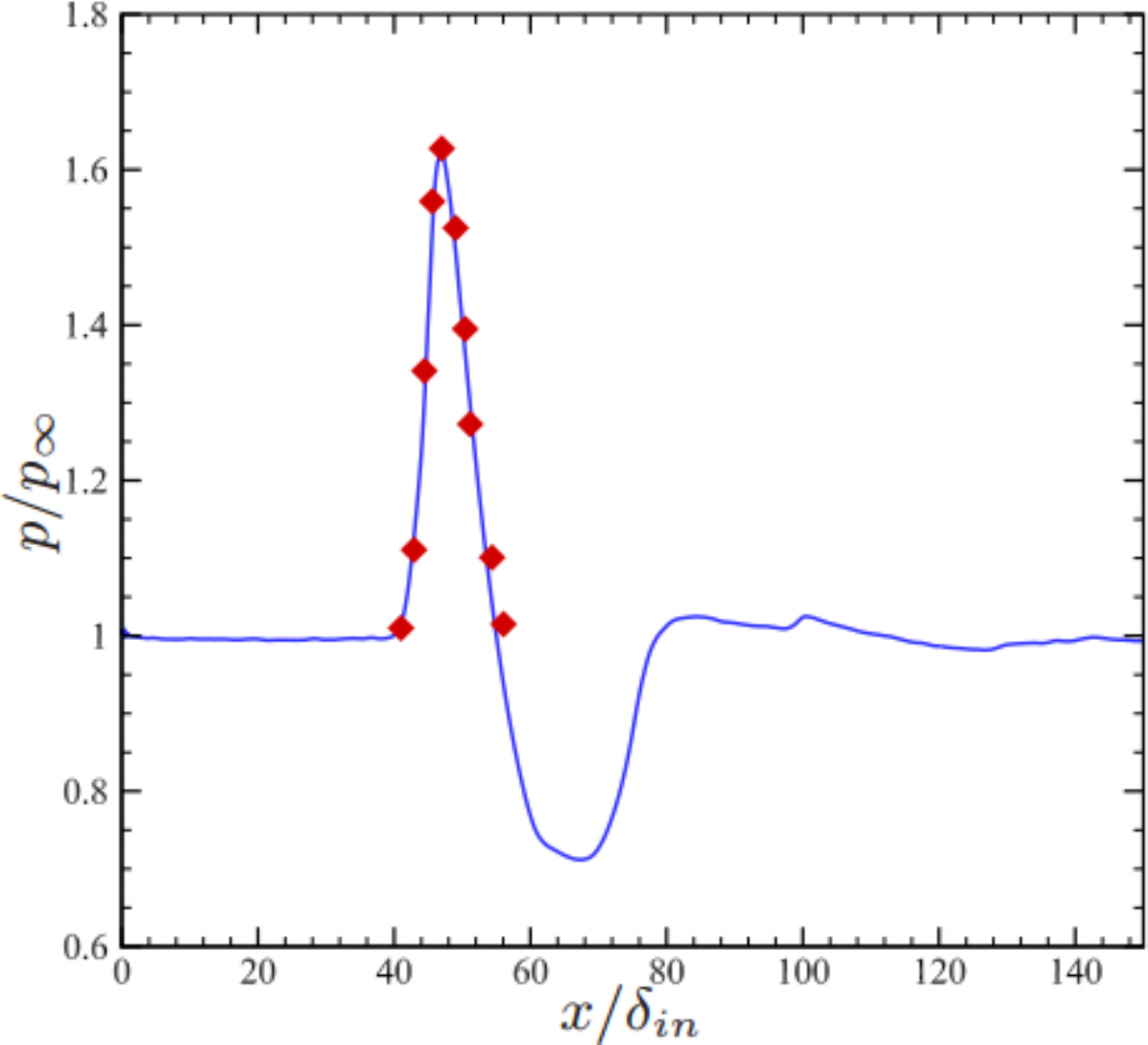}
        \\ \vskip1.em
	\caption{Mean wall pressure profiles at different spanwise positions:
	  (a), $z/\delta_{in} = 0$; (b), $z/\delta_{in} = 4.35$.
	  Lines, DNS; symbols, experiment~\citep{Hale2015-Interaction}.}
	\label{fig:p_comp}
\end{figure}

An important feature in shock/boundary layer interactions is the steady and unsteady wall pressure load,
which may have important impact on the behavior of the underlying structural components.
The mean wall pressure ${\bar p_w}$ across the interaction zone is compared in figure~\ref{fig:PSP}
with the experimental data of~\citet{Hale2015-Interaction}, obtained with  pressure-sensitive paint.
It should be noted that in the experimental arrangement the cone is held to place 
by a thick cylindrical sting, which may affect the flow in its wake and at the wall. 
Despite this difference and the previously cited Reynolds number disparity, 
figure~\ref{fig:PSP} shows nevertheless
favourable comparison, at least limited to the experimental measurement window.
This is better seen in figure~\ref{fig:p_comp}, where we show mean pressure profiles at two spanwise locations.
The figure well brings out the N-wave wall signature of the conical shock, with pressure rising sharply at the 
nominal shock impingement location, then decreasing almost linearly to a value which is lower than the free-stream,
and increasing again upon impingement of the recompression shock.
The peak value along the symmetry line is attained at approximately $ x/{\delta _{in}} \approx 46 $,
whereas the peak is slightly shifted downstream at the other spanwise station, 
owing to the hyperbolic shape of the shock foot.
After passing through the shock, pressure decreases upon passage of the expansion
waves generated at the cone trailing edge. As flow develops downstream,
pressure rises again in the second APG region ($ x/{\delta _{in}} \approx 66$).
Finally, the flow returns to a nearly ZPG
condition. 
It should be noted that the same wall pressure pattern was observed in the experiments of \citet{Gai2000-Interaction}.
Based on the comparison reported in figure~\ref{fig:p_comp}, we can confidently
conclude that the DNS results adequately reproduce the flow physics.

\begin{figure}
	\centering
	\includegraphics[width=10.0cm]{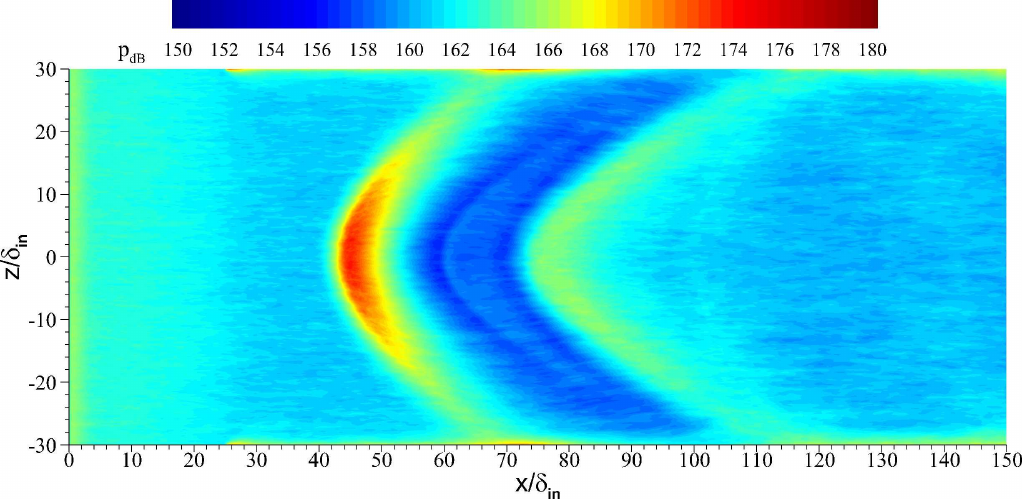}
	\caption{Contours of root-mean-square pressure fluctuations ($p_{rms}$)
	in the wall plane, in $dB$ scale, $p_{dB} = 20 \log_{10} (p / 2 \cdot 10 ^{-5} \mathrm{Pa})$, assuming
        $p_{\infty} = 1$atm. Contour levels are shown from $150$ to $180$, from blue to red.}
	\label{fig:wall_prms}
\end{figure}

The root-mean-square pressure fluctuations contours are shown in figure~\ref{fig:wall_prms}. 
Strong spatial connection of this distribution is found with that of the vortical
structures shown in figure~\ref{fig:3d_vortex}.
In particular, the largest values of $p_{rms}$ are found at the interacting shock foot,
especially around the symmetry axis where the shock is stronger.
Consistent decrease of the fluctuating pressure loads is observed past the incident shock
in the FPG region, which is depleted with eddies.
A secondary peak is observed further downstream, corresponding to re-formation
of the vortical structures.

\subsection{Boundary layer development}

\begin{figure}
	\centering
	\psfrag{x}[][][1.0]{$ x/\delta_{in} $}
	\psfrag{a}[][][1.0]{$ \delta^*/\delta_{in}, \theta/\delta_{in} $}
	\psfrag{b}[][][1.0]{$ H_i $}
    \includegraphics[width=10.0cm,clip]{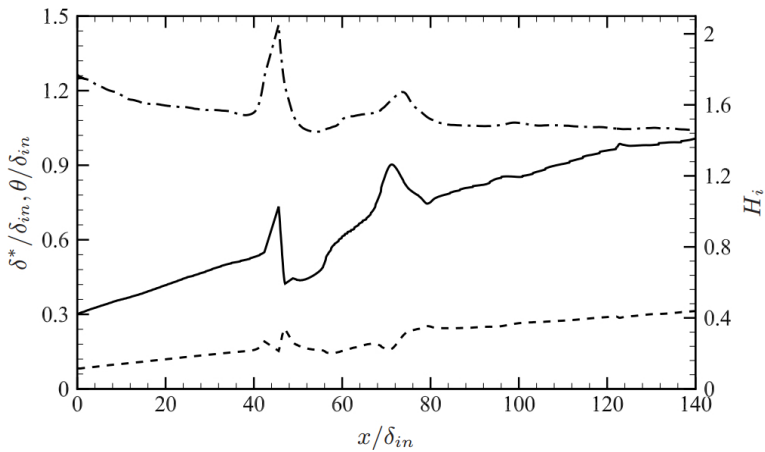}
	\caption{Streamwise evolution of boundary layer properties in the symmetry plane:
	displacement thickness (solid), (b) momentum thickness (dashed), incompressible shape factor (dash-dot).}
	\label{fig:bl_evolution}
\end{figure}

In order to characterize the spatial development of the wall boundary layer,
we preliminarily analyse the distributions of the displacement and momentum thicknesses,
defined in equations \eqref{eqn:dstar}, \eqref{eqn:theta}.
The streamwise distributions of the displacement thickness ($\delta ^*$),
of the momentum thickness ($\theta$)
and of the incompressible shape factor ($H_i = \delta_i^* / \theta _i$)
along the symmetry plane are shown in figure~\ref{fig:bl_evolution}.
Upstream of the cone leading edge $H_i$ varies between 1.55 and 1.75,
as appropriate for canonical ZPG boundary layers at low Reynolds number~\citep{Wu-2009-Direct}. 
Corresponding to the leading edge of the interaction zone, 
the displacement thickness increases sharply,
whereas the momentum thickness is relatively unaffected.
Hence, the shape factor attains a maximum value $H_i \approx 2.0$,
indicating a less full velocity profile.
The displacement thickness undergoes a sudden drop 
in the FPG region followed by gradual recovery, 
and a new peak is observed in the second APG region, also corresponding to a peak of $H_i$. 
Past $x/\delta_{in}\approx 90$, the boundary layer recovers an equilibrium state
which is an ideal continuation of the undisturbed upstream state.

\begin{figure}
	\psfrag{x}[][][1.2]{$ x/\delta_{in} $}
    \psfrag{y}[][][1.2]{$ \beta $}
%   \psfrag{a}[][][1.0]{$ 39 $}
%   \psfrag{b}[][][1.0]{$ 46 $}
%   \psfrag{c}[][][1.0]{$ 66 $}
%   \psfrag{d}[][][1.0]{$ 80 $}
    \psfrag{a}[][][1.0]{}
    \psfrag{b}[][][1.0]{}
    \psfrag{c}[][][1.0]{}
    \psfrag{d}[][][1.0]{}
    \psfrag{e}[][][0.8]{ZPG1}
    \psfrag{f}[][][0.8]{APG1}
    \psfrag{k}[][][0.8]{FPG}
    \psfrag{h}[][][0.8]{APG2}
    \psfrag{i}[][][0.8]{ZPG2}
	\centering
	\includegraphics[width=10.0cm]{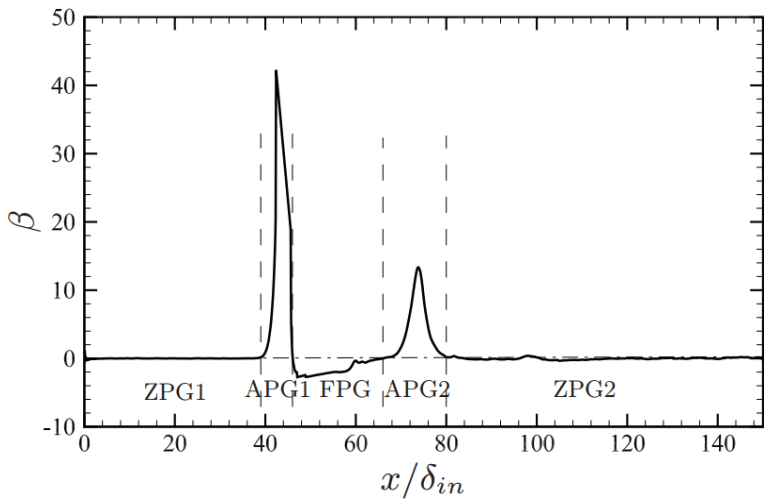}
	\caption{Streamwise distribution of Clauser pressure gradient parameter 
	in the symmetry plane.}
	\label{fig:Clauser}
\end{figure}

Non-equilibrium states of boundary layers upon imposed pressure gradient are traditionally
analyzed in terms of Clauser pressure gradient parameter, defined as~\citep{Clauser1954-Turbulent}
\begin{equation}
 \beta  = \frac{\delta^*}{\rho_w u_\tau^2} \frac{\diff {\bar p}_w}{\diff x} ,
\label{eqn:beta}
\end{equation}
whose distribution along the symmetry axis is shown in figure~\ref{fig:Clauser}.
According to the DNS data, the flow field may be divided into five parts:
ZPG1, the upstream ZPG region, with $\beta \approx 0$;
APG1, the first APG region, where $\beta$ exhibits a sharp positive peak;
FPG , where $\beta$ is negative as the flow accelerates; 
APG2, the second APG region, where $\beta$ attains a second positive peak;
ZPG2, the downstream ZPG region where equilibrium conditions are recovered.

\begin{figure}
	\psfrag{y}[][][1.2]{$u _e / u_\infty$}
	\psfrag{x}[][][1.2]{$\delta^* / \delta_{in}$}
	\psfrag{a}[][][1]{$\Lambda=-1.92$}
	\psfrag{b}[b][][1]{FPG}
	\psfrag{c}[][][1]{ }
	\psfrag{d}[tr][br][1]{APG2}
	\psfrag{e}[bl][][1]{$\Lambda=0.22$}
	\psfrag{f}[t][b][1]{APG1}
	\psfrag{g}[][][1]{ }
	\psfrag{h}[l][][1]{$\Lambda=0.22$}
	\centering
        \includegraphics[width=10.0cm,angle=0,clip]{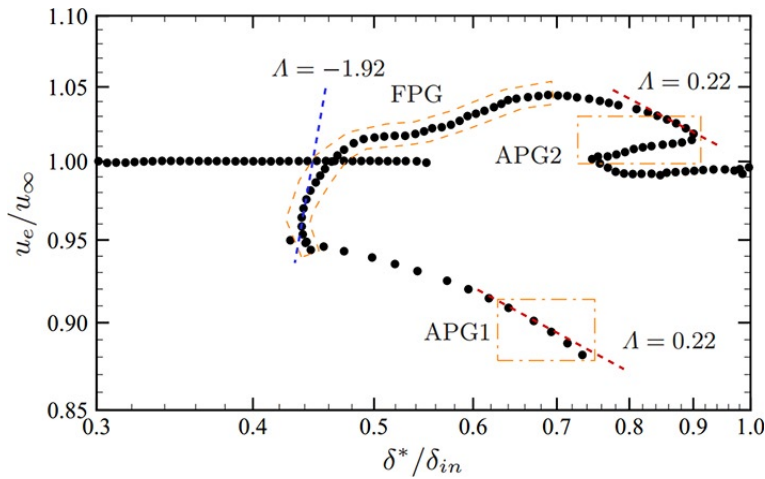}
	\caption{External velocity as a function of boundary layer displacement thickness along the
        symmetry plane. Dashed lines denote the power-law scalings given in equation~\eqref{eqn:delta-u-castillo}.}
	\label{fig:lambda}
\end{figure}

\citet{Castillo2001-Similarity} proposed that the proper velocity scale for the 
outer part of the boundary layer is the external velocity $u_e$ rather than the friction velocity, and
accordingly they introduced a modified pressure gradient parameter defined as
\begin{equation}
\Lambda  =
\frac{\delta}{\rho_e u_e^2 \diff \delta / \diff x} \frac{\diff {\bar p}_w}{\diff x} =
- \frac{\delta}{u_e \diff \delta / \diff x} \frac{\diff u_e}{\diff x} ,
\label{eqn:castillo}
\end{equation}
where $\delta$ is any measure of the boundary layer thickness.
They suggested that, at least in the absence of mean flow separation, the admissible
equilibrium states corresponding to self-similar velocity distributions 
must have constant value of $\Lambda$, 
which in turn implies power-law dependence of the boundary-layer thickness on the external velocity, namely
\begin{equation}
\delta \sim u_e^{-1/\Lambda} .
\label{eqn:delta-u-castillo}
\end{equation}
It was found that most experimental data for boundary layers in adverse and favourable 
pressure gradient are in fact in equilibrium according to this definition, 
and three values of $\Lambda$ were found to be possible: $\Lambda = 0.22$ in APG, 
$ \Lambda = -1.92 $ in FPG and 
$ \Lambda = 0 $ in ZPG. 
This is scrutinized in figure~\ref{fig:lambda}, where we show 
the boundary layer external velocity (evaluated at $y=\delta_e$) as 
a function of the boundary layer displacement thickness.
The figure suggests very partial success of theoretical predictions,
mainly limited to the APG regions, where scaling not far from the predicted $\Lambda = 0.22$ is observed,
whereas the FPG region seems to have a behavior very different from the theoretical equilibrium state.

\begin{figure}
   	\centering
	\psfrag{a}[][][1.0]{$ ({u_{v{d_e}}} - {u_{vd}})/{u_{vd}}_e\cdot\delta /{\delta ^*} $}
	\psfrag{y}[][][1.0]{}
	(a)\includegraphics[width=6.25cm]{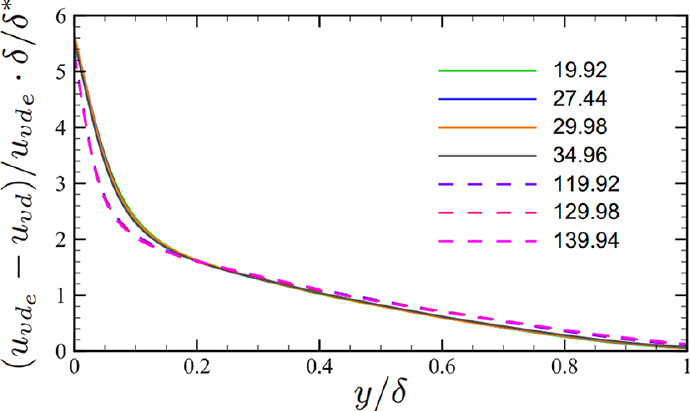}
	\psfrag{a}[][][1.0]{}
	(b)\includegraphics[width=6.25cm]{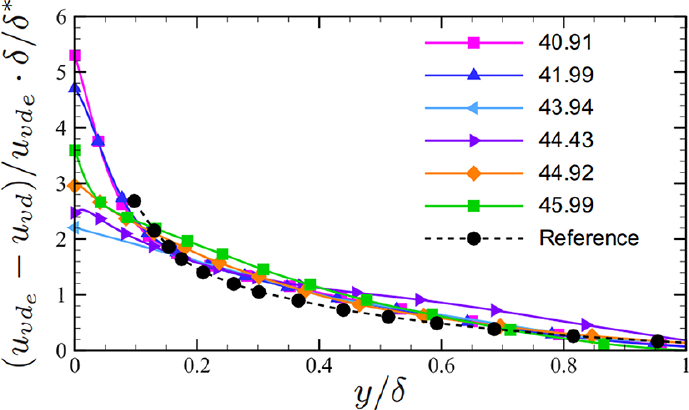} \\ \vskip1.em
	\psfrag{y}[][][1.0]{$ y/ \delta $}
	\psfrag{a}[][][1.0]{$ ({u_{v{d_e}}} - {u_{vd}})/{u_{vd}}_e\cdot\delta /{\delta ^*} $}
	(c)\includegraphics[width=6.25cm]{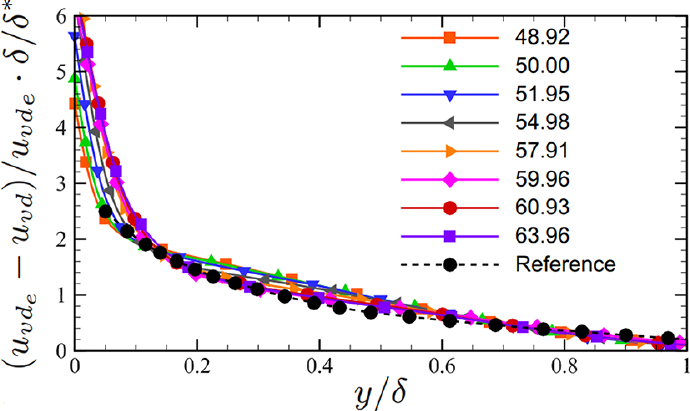}
	\psfrag{a}[][][1.0]{}
	(d)\includegraphics[width=6.25cm]{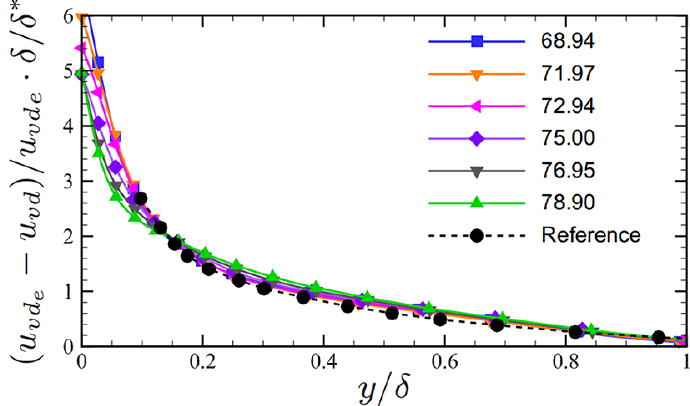}  \\ \vskip1.em
	\caption{Mean velocity defect profiles at different streamwise stations 
        in region ZPG1+ZPG2 (a), APG1 (b), FPG (c), APG2 (d).
	Reference data from \citet{Song-2004-Reynolds} and \citet{Aubertine-2005-turbulence} are shown in panels (c) 
        and (b, d), respectively.}
	\label{fig:zagarola}
\end{figure}

Self-similarity of the velocity profiles in the various zones is checked in figure~\ref{fig:zagarola},
where we show the van Driest transformed velocity multiplied by $\delta/\delta^*$ to 
remove effects of upstream conditions and local Reynolds number on the outer velocity profile~\citep{Zagarola-1998-Mean}.
In the ZPG1 and ZPG2 regions the velocity profiles do in fact collapse on a single curve, 
with the obvious exception of the near-wall region which suffers Reynolds number dependence.
Similar considerations can be made for the FPG and the APG2 regions, which show a good 
degree of universality, and in which the defect velocity profiles 
well match reference experimental data for FPG flows~\citep{Song-2004-Reynolds} 
and for APG flows~\citep{Aubertine-2005-turbulence}.
Not surprisingly, large deviations from a common distribution are observed in APG1 region,
which experiences strong APG conditions and in which the flow even undergoes mean separation.

\subsection{Analysis of flow reversal}

\begin{figure}
	\psfrag{x}[][][1.2]{$x / \delta_{in}$}
	\psfrag{y}[][][1.2]{$C_f$}
	\centering
	\includegraphics[width=10.0cm]{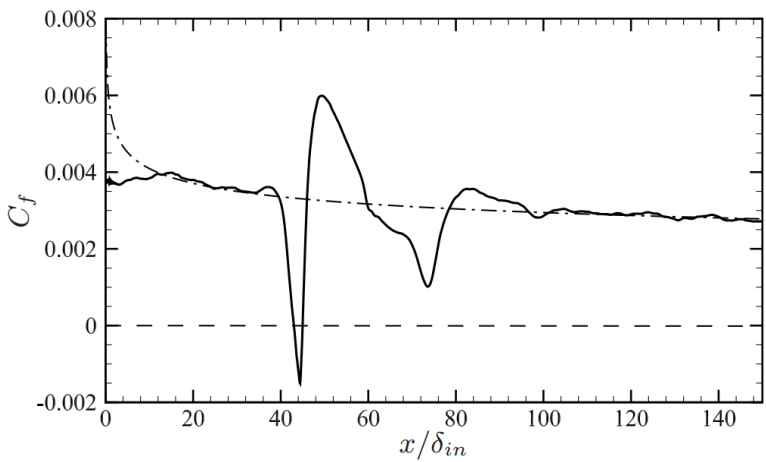}
	\caption{Distribution of mean skin friction coefficient in the symmetry plane.
	The dash-dotted line refers to equation~\eqref{eq:Cf}.}
	\label{fig:cf}
\end{figure}

The imposed adverse pressure gradient causes flow retardation along with locally reversed flow.
The distribution of the mean friction coefficient $C_f = 2 \tau_w / \rho_w u_{\infty}^2$, 
with $\tau_w = \mu_w (\partial{\tilde{u}} / \partial y)_w$ is shown in the symmetry plane in figure~\ref{fig:cf}. 
Upstream of the interaction zone and in the downstream ZPG2 region the friction coefficient well follows
the power-law behavior predicted by simple theory~\citep{Smits-2006-Turbulent}, 
namely
\begin{equation}
C_f = k \, \Rey_x^{-1/7}, \label{eq:Cf}
\end{equation}
with $k = 0.0192$, with the obvious exception of the inflow region  
where the boundary layer is not properly developed yet.
Mean flow reversal in the symmetry plane is observed in the APG1 region, whereas $C_f$ overshoots
its upstream value in the FPG region right past the reattachment point. 
A secondary minimum of $C_f$ is found in the APG2 region, where however the flow 
is not detached in mean sense.

\begin{figure}
	\centering
        (a)
	\includegraphics[width=12.0cm]{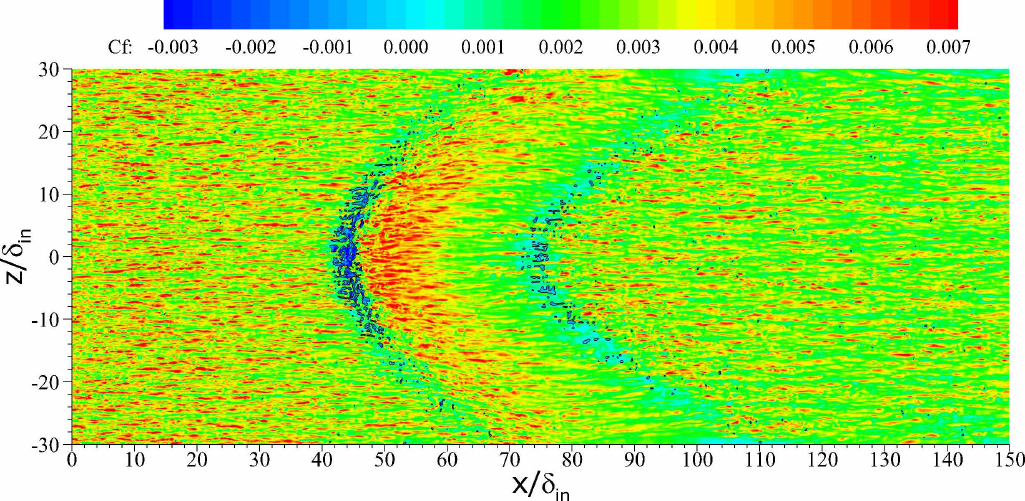} \\
        (b)
	\includegraphics[width=12.0cm]{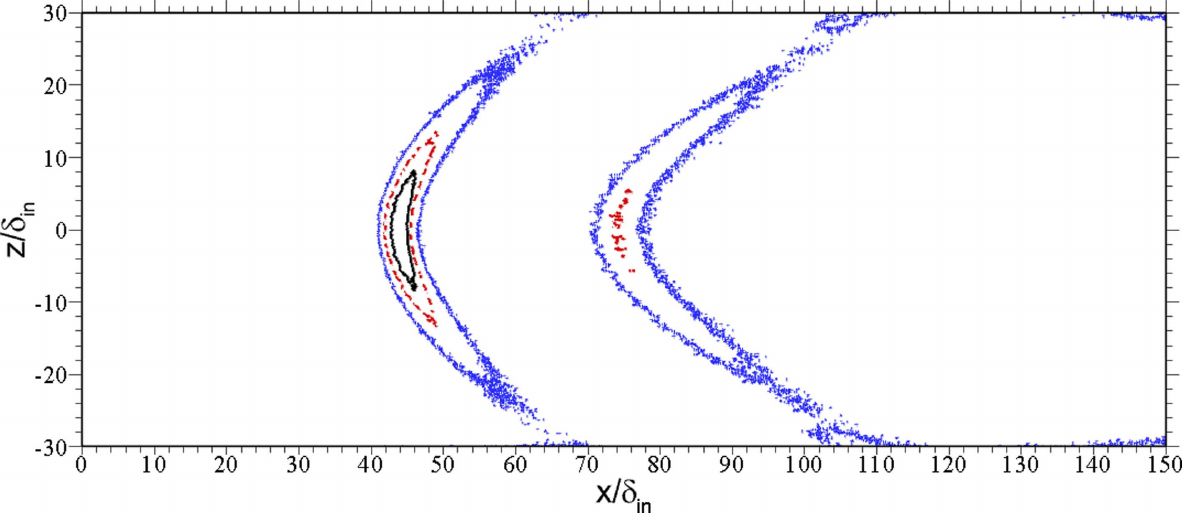} 
	\caption{Instantaneous contours of wall friction (a) and of flow reversal probability $\overline{\gamma}$ (b) at the wall surface.
	The contour levels $\overline{\gamma}=$ 0.5 (solid), 0.2 (dashed), 0.01 (dotted) are shown in panel (b).}
	\label{fig:separation_prob}
\end{figure}

Additional insight into the nature of wall flow reversal 
may be gained from figure~\ref{fig:separation_prob},
where we show the instantaneous wall friction and the 
statistical frequency of events with wall flow reversal (say $\overline{\gamma}$).
Figure~\ref{fig:separation_prob}(a) shows that both the APG1 and APG2 regions 
feature a substantial fraction of flow reversal events, which are however scattered 
and interspersed with regions of attached flow. On the other hand, the FPG region
is depleted with reverse flow events, and it features friction excess over the upstream value.
The intermittency data are shown in panel (b).
The data can be interpreted  in light of
the classification proposed by \citet{Simpson1989-Turbulent}, 
according to which incipient detachment occurs with instantaneous backflow 1\% of the time; 
intermittent transitory detachment occurs with instantaneous backflow 20\% of the time; 
transitory detachment (amounting to mean flow reversal) happens with 50\%
probability of instantaneous backflow. 
According to figure~\ref{fig:separation_prob}(b),
mean flow reversal is found to occur only in the APG1 region around the symmetry line, whereas
the flow becomes unidirectional in the mean away from the symmetry axis where pressure gradient is milder.
Hence, the reversed flow region tends to become narrower along the $z$ direction, eventually vanishing at $z \approx \pm 7.7 \delta_{in}$.
Transitory detachment is observed both in the APG1 and APG2 regions, 
to a much lesser extent in the latter case.
Interestingly, an extended region with incipient detachment is observed in both APG zones,
with no significant size difference.

\begin{figure}
	\centering
        (a)
	\psfrag{X}[t][][1.2]{$x / \delta_{in}$}
	\psfrag{Y}[][][1.2]{$z / \delta_{in}$}
%	\psfrag{S}[][][1.4]{\color{blue}{S}}
%	\psfrag{N}[][][1.4]{\color{red}{N}}
%	\psfrag{F}[][][1.4]{\color{green}{F}}
	\includegraphics[width=5.0cm,angle=0,clip]{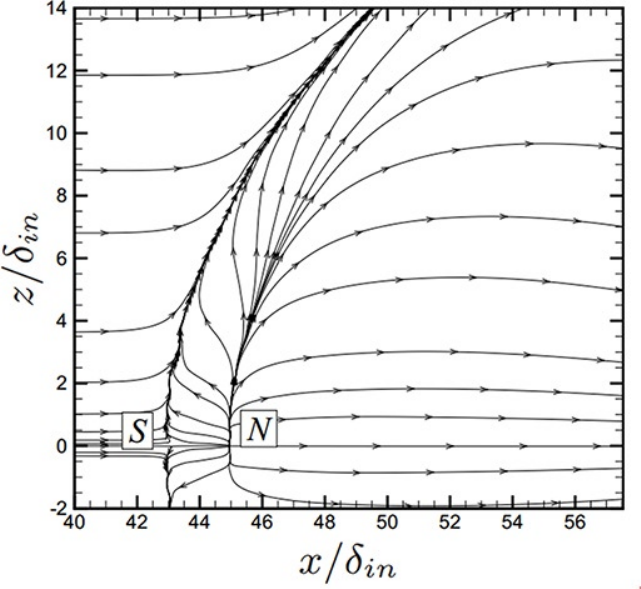} \hskip1.em
        (b)
	\psfrag{X}[t][][1.2]{$x / \delta_{in}$}
	\psfrag{Y}[b][][1.2]{$y / \delta_{in}$}
	\includegraphics[width=5.0cm,angle=0,clip]{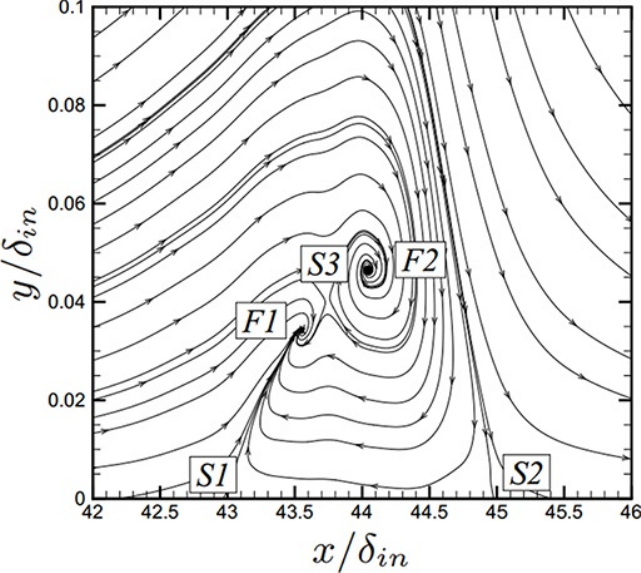}
        \vskip1.em
	\caption{Limit wall streamlines of mean flow field (a),
        and streamtraces in the symmetry plane (b), highlighting critical points:
        N, nodes; S, saddle points; F, foci. Note that axes in panel (b) are not to scale.}
	\label{fig:streamlines}
\end{figure}

The flow topology is further brought out from the analysis of the limit streamtraces in 
the wall plane and of the streamtraces in the symmetry plane,
reported in figure~\ref{fig:streamlines}. 
The wall limiting streamlines exhibit a pattern similar to that resulting from separation 
induced by an obstacle~\citep{delery_01}, and the flow features a saddle point 
associated with three-dimensional flow separation. The flow past the separation streamline 
is characterized by the presence of a node at $x/\delta_{in} \approx 45$ associated 
with reattachment in the symmetry plane. The wall streamtraces are issued from the nodal point 
and partly converge into the separation line and partly are deflected downstream.
This wall pattern is classically associated with the formation of a horseshoe vortex bending in the
downstream direction, as is clarified in figure~\ref{fig:streamlines}(b).
A very similar organization was also recovered in the experiments of \citet{Gai2000-Interaction}.
Visualization in the symmetry plane highlights the presence of two saddle points at the wall (S1 and S2),
being respectively the signature of the separation and reattachment point, and two foci of stretching type (F1 and F2),
separated by a third saddle point (S3). This is to indicate that the horseshoe vortex is in fact
split into two branches, which merge together moving away from the symmetry plane
(not shown). It should be noted that the region of reversed flow, which may be defined as the region 
between the wall and the streamtrace entering saddle point $S_2$, is quite shallow, extending
up to $y_R/\delta_{in} \approx 0.11$. Hence the aspect ratio of the separation bubble is very small,
being $\AR = y_R/(x_{S_2}-x_{S_1}) \approx 0.057$, which is a typical value for turbulent 
separation bubbles, but much less than laminar ones~\citep{kiya_83}.

\subsection{Turbulence statistics}

\begin{figure}
  \centering
  (a)
  \psfrag{x}[t][][1.2]{$x / \delta_{in}$}
  \psfrag{y}[b][][1.2]{$y / \delta_{in}$}
  \includegraphics[width=10.5cm,angle=0,clip]{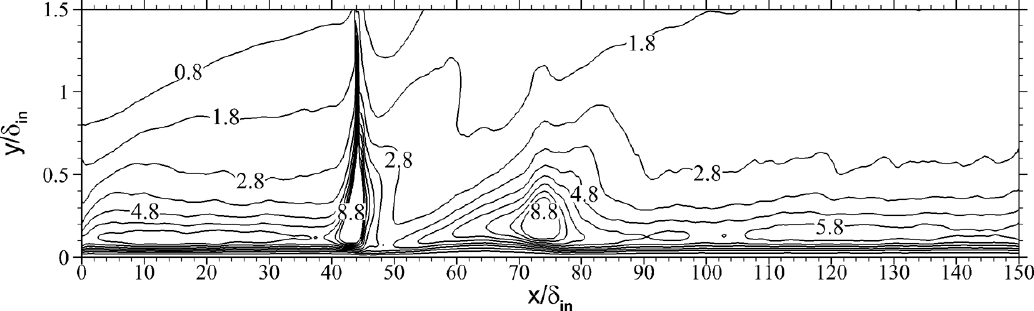} \\ \vskip1.em
  (b)
  \includegraphics[width=10.5cm,angle=0,clip]{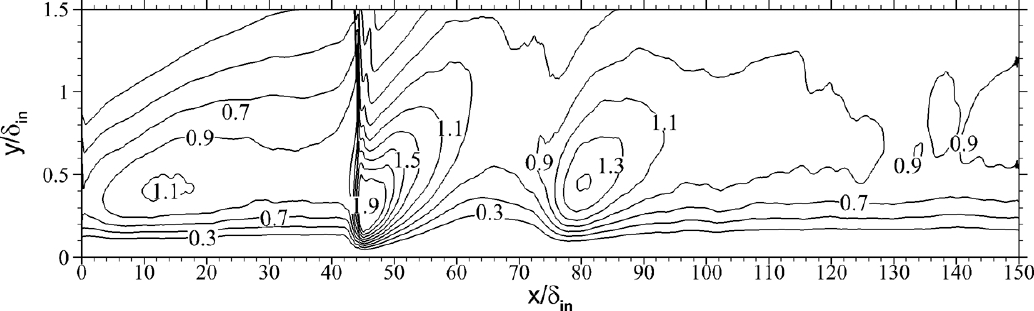} \\ \vskip1.em 
  (c)
  \includegraphics[width=10.5cm,angle=0,clip]{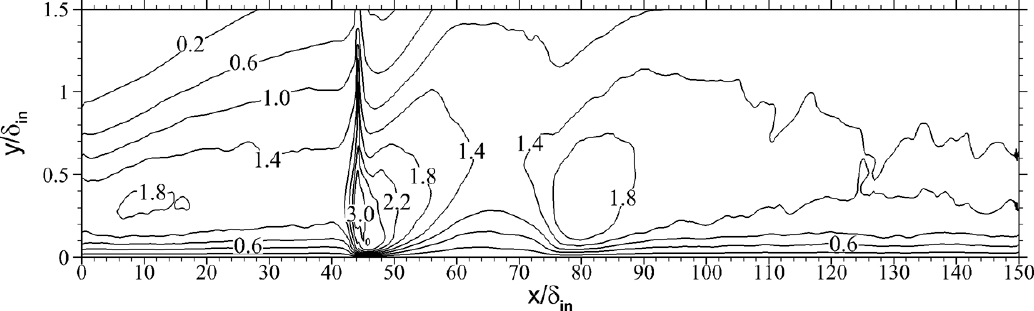} \\ \vskip1.em
  (d)
  \includegraphics[width=10.5cm,angle=0,clip]{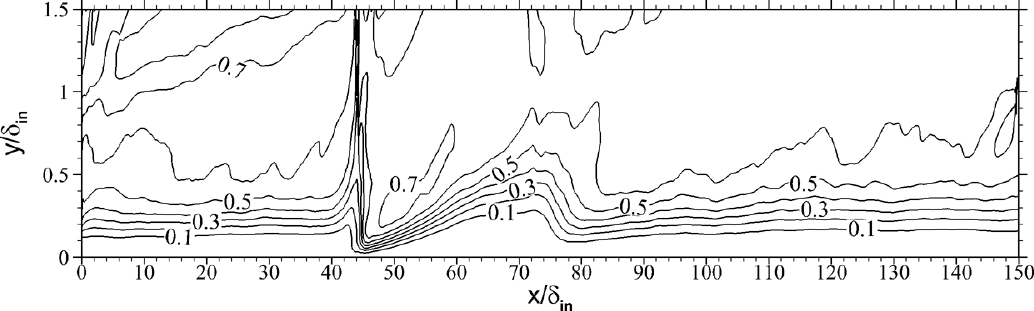} \\ \vskip1.em
  \caption{Maps of density-scaled velocity correlations ($\overline{\rho}/\rho_w \widetilde{u''_i u''_j}$) in the symmetry plane, 
  scaled by the friction velocity at the upstream reference station (a, $i=j=1$; b, $i=j=2$; c, $i=j=3$) and anisotropy indicator (d), as defined in equation~\eqref{eq:lumley}.}
  \label{fig:Restress}
\end{figure}

The structural modifications of boundary layer turbulence upon interaction with the conical shock are analyzed in this section. 
For that purpose, contours of the density-scaled velocity correlations $\overline{\rho}/\rho_w \widetilde{u''_i u''_j}$ 
in the symmetry plane are shown in figure~\ref{fig:Restress}, upon normalization by the friction velocity at the upstream 
reference station ${u_{\tau}}_{ref}$.
The figure supports general amplification of all turbulence intensities across the APG zones, however to a different extent.
Specifically, the longitudinal velocity variance is amplified (with respect to its upstream value) by factor of about 2 
through the APG1 zone, and by 1.6 in APG2. Similar amplifications are found for the wall-normal velocity correlation, whereas the spanwise 
velocity variance increases by a factor of $1.8$ across APG1, and by a factor $1.1$ in APG2.
As expected, peaks of the normal turbulent stresses in the ZPG zones are found to lie closer to the wall 
for the streamwise component, and further away for the other components. 
To better highlight modifications of the turbulent stress tensor, in figure~\ref{fig:Restress} we show the map
of the Reynolds stress anisotropy function~\citep{lumley_78}, defined as
\begin{equation} 
F=1+9 II + 27 III, \label{eq:lumley}
\end{equation} 
where $II$ and $III$ are the invariants of the anisotropy stress tensor, 
$b_{ij} = \tau^*_{ij} / \tau^*_{ii} - 1/3 \delta_{ij}$,
and which is a measure of the approach to either two-component turbulence 
(corresponding to $F =0$) or a three-component isotropic state (corresponding to $F=1$). 
Consistent with previous DNS data, figure~\ref{fig:Restress} shows that in ZPG regions $F$ is 
increasing from a nearly zero value at the wall where the flow is dominated by 
streaks, to a nearly uniform value of about $0.6$ in the outer part of the boundary layer.
As the flow crosses the interacting shock in the APG1 region the anisotropy indicator
increases at any given $y$, indicating approach of turbulence to an isotropic state. 
The opposite effect is observed in the FPG region which shows consistent decrease of this
indicator, accompanied by the previously noted re-formation of the streaks, and again
increase is observed in APG2 region, prior to return to an equilibrium state.

\section{Conclusions}\label{sec:conclusions}

The interaction of a conical shock wave with a turbulent
boundary layer at free-stream Mach number $ {M_\infty } = 2.05 $,
half-cone angle $ {\theta _c} = {25^ \circ } $ and Reynolds number
$ \Rey_\theta \approx 630 $ has been analysed by means of DNS
of the compressible Navier-Stokes equations. 
Detailed flow statistics have been presented, including mean flow properties and turbulent fluctuations.
Particular effort has been made to characterize the geometry of the shock system and the
three-dimensional features of the interaction region.

Consistent with experimental observations, 
the mean flow pattern is found to include of a main conical shock which imposes a hyperbolic footprint
on the underlying flat plate, and which causes 
thickening and local separation of the developing boundary layer. The incident shock is reflected 
as two conical shock waves, one arising because of the
upstream influence mechanisms, and the second past the boundary layer reattachment.
As theoretically predicted, analysis of the entropy fields obtained from DNS allows to discern
transition from regular to Mach reflection with a distinct shock stem moving away from the symmetry plane,
although the transition point is probably much farther than suggested by the inviscid theory.
The compression waves originating within the cone wake also tend to coalesce to form a secondary weaker conical shock,
which interacts with the boundary layer further downstream.
Overall, this wave pattern imparts a distinctive N-wave signature at the wall, 
with a primary APG region followed by a FPG region, which is then closed by a secondary APG region
bringing the boundary layer back to an equilibrium state.
Very good agreement of the computed wall pressure signature is found with respect to reference 
experimental measurements.

The imposed wall pressure gradient is responsible for strong turbulence non-equilibrium in the boundary layer.
In this respect, flow visualization in the upstream ZPG region show that the outer part of the boundary layer is populated by
hairpin-shaped vortices as well as more asymmetric, cane-shaped vortices.
Vortices tend to disappear in the FPG region, and to reform past the recompression shock.
Correspondingly, the near-wall streaks of high- and low-speed fluid which are present in the incoming ZPG region, 
are suppressed in the APG regions, to quickly reform downstream.
Statistical analysis of flow reversal zones highlights
mean flow separation in the primary APG zone accompanied with formation of 
a horseshoe vortex, whereas the secondary APG zone is characterized by
intermittent detachment with scattered spots of instantaneous flow reversal.

From a quantitative standpoint we find that turbulence non-equilibrium is only partially captured 
by existing theoretical frameworks. In particular, we find that self-similarity 
proposed by~\citet{Castillo2001-Similarity} is partly attained in APG zones, whereas 
large departures are found in the FPG zone.
Different amplification of the Reynolds stress components is observed
in the APG regions, accompanied by a change in the geometry of the Reynolds stress tensor.
Specifically, isotropy is favoured in APG regions, whereas a anisotropic two-component state is recovered 
in the FPG zone, associated with reformation of the streaks.

Further studies will be devoted to enlarging the DNS database and include cases with different shock 
strengths, and to further characterize the unsteady wall signature. Comparison of simplified RANS 
predictions with DNS would also be of great interest.

We acknowledge that the results reported in this paper have been achieved using the PRACE Research Infrastructure
resource MARCONI based at CINECA, Casalecchio di Reno, Italy.
The China Scholarship Council is gratefully acknowledged for supporting
Fengyuan Zuo (No.201706830060) as a joint Ph.D.~student at Sapienza University of Rome.

\bibliographystyle{jfm}

\bibliography{Ref-CSBLI1}

\end{document}